%% file: WVU_MDC2.tex
\documentclass{iopart}
\usepackage{iopams}
\usepackage{color}
\usepackage{graphicx}
\usepackage{enumitem}
\usepackage{xspace}
\usepackage{multirow}
\usepackage{hyperref}
\usepackage{natbib}
\usepackage[caption=false]{subfig}

\def\be{\begin{equation}}
\def\ee{\end{equation}}
\def\ba{\begin{align}}
\def\ea{\end{align}}

\newcommand{\enterprise}{{\texttt{enter\-prise}}\xspace}
\newcommand{\libstempo}{{\texttt{libstempo}}\xspace}
\newcommand{\ptmcmc}{{\texttt{PTMCMC\-Sampler}}\xspace}
\newcommand{\bayesephem}{{\texttt{Bayes\-Ephem}}\xspace}

\newcommand{\red}{{\mathrm{RN}}}
\newcommand{\yr}{{\mathrm{yr}}}
\newcommand{\dd}{{\mathrm{d}}}

\newcommand{\GWB}{{\mathrm{GWB}}}

\begin{document}

\title[IPTA MDC2 Results]{Results for the International Pulsar Timing Array Second Mock Data Challenge: New Techniques and Challenges for the Detection of Low-Frequency Gravitational-Wave Signals}

\author{P.\,T.\,Baker$^{1,2}$, P.\,R.\,Brook$^{2,3}$, W.\,C.\,Fiore$^{2,3}$, N.\,Garver-Daniels$^{2,3}$, A.\,R.\,Kaiser$^{2,3}$, M.\,T.\,Lam$^{5,2,3}$, B.\,J.\,Shapiro-Albert$^{2,3}$, C.\,A.\,Witt$^{2,3}$}

\address{$^1$Department of Physics and Astronomy, Widener University, One University Place, Chester, PA 19013, USA}
\address{$^2$Center for Gravitational Waves and Cosmology, West Virginia University, Chestnut Ridge Research Building, Morgantown, WV 26505, USA}
\address{$^3$Department of Physics and Astronomy, West Virginia University, P.O. Box 6315, Morgantown, WV 26506, USA}
\address{$^4$School of Physics and Astronomy, Rochester Institute of Technology, Rochester, NY 14623, USA}
\ead{paul.baker@nanograv.org}


\begin{abstract}

We present a detailed analysis of the International Pulsar Timing Array (IPTA) Second Mock Data Challenge.
We tested our analysis methods using the open datasets, and then analyzed the closed datasets.
In both the open and the closed datasets, we were able to detect some, but not all, of the injected gravitational wave signals.
This work presents two search cases that are not well explored in the pulsar timing array (PTA) literature:
a simultaneous search for a stochastic GW background and an individual loud super-massive black hole binary (SMBHB)
and a simultaneous search for two SMBHB sources.
While we have constructed a cohesive framework for performing these GW searches, our analyses required fine-tuning of the sampling method used in order to appropriately converge.
Given the nature of real PTA data in which multiple sources will be present in data, improved techniques will be required in the future to accurately detect and characterize these GW signals.

\end{abstract}


\submitto{\CQG}

\section{Introduction}
\label{sec:introduction}
\input{sec_intro.tex}

\section{The IPTA Mock Data Challenge}
\label{sec:mdc}
\input{sec_mdc.tex}

\section{Data Analysis with \enterprise}
\label{sec:analysis}
\input{sec_analysis.tex}

\section{Verification of Open Datasets}
\label{sec:open}
\input{sec_open.tex}

\section{Results on Closed Datasets}
\label{sec:closed}
\input{sec_closed}

\section{Discussion} 
\label{sec:discuss}
\input{sec_discuss.tex}

\ack
\textit{Author Contributions.} This work is the product of the entire team and we list specific contributions of individual authors below.
P.T.B. coordinated the group effort and provided direction for the team.
C.A.W. developed methods in \enterprise to search for GWB + CW and 2xCW signals, and to sample over CW frequency, implemented these methods in the respective datasets, and conducted normal CW analyses for all datasets.
P.R.B., W.C.F., A.R.K., B.J.S., and C.A.W. conducted pulsar noise and GWB analyses.
P.T.B., P.R.B., W.C.F., A.R.K., M.T.L., B.J.S., and C.A.W. contributed to the interpretation of results and paper writing.
N.G. assisted with managing computational resources.
\par
\textit{Acknowledgments.} The authors would like to thank Jeffery S.~Hazboun, who led the effort to organize and release MDC2, for promptly answering our questions and indulging in useful discussions.
The NANOGrav Project receives support from NSF Physics Frontiers Center award number 1430284. 
B.J.S. and W.C.F. acknowledge support from West Virginia University through the STEM Mountains of Excellence Graduate Fellowship.
C.A.W. acknowledges support from West Virginia University through the Outstanding Merit Fellowship for Continuing Doctoral Students, and is also supported by NSF awards \#1458952 and \#1815664. 
P.R.B. is supported by Track I award OIA-1458952.
We are grateful for computational resources provided by the Leonard E. Parker Center for Gravitation, Cosmology and Astrophysics at the University
of Wisconsin-Milwaukee, which is supported by NSF Grants
0923409 and 1626190.
This research made use of the Super Computing System (Spruce Knob) at WVU, which
is funded in part by the National Science Foundation EPSCoR Research Infrastructure Improvement Cooperative Agreement \#1003907, the state of West Virginia (WVEPSCoR via the Higher Education Policy Commission) and WVU. We acknowledge use of Thorny Flat at WVU, which is funded in part by the National Science Foundation Major Research Instrumentation Program (MRI) Award \#1726534 and WVU.
\par
\textit{Software.}
\enterprise \cite[][\url{https://github.com/nanograv/enterprise}]{enterprise},
\libstempo \cite[][\url{https://github.com/vallis/libstempo}]{libstempo},
\ptmcmc \cite[][\url{https://github.com/jellis18/ptmcmcsampler}]{ptmcmc}

\appendix
\input{sec_appendix.tex}

\bibliographystyle{iopart-num}
\clearpage
\input{biblio}

\end{document}

%% file: sec_intro.tex
The detection of low-frequency gravitational waves (GWs) by a pulsar timing array (PTA) is imminent \citep[e.g.][]{Taylor+2016}.
Unlike recent measurements by the Laser Interferometer Gravitational Wave Observatory (LIGO) in which one can estimate the noise properties of the detector from times when GW signals are not present \citep[e.g.][]{Was+2010}, GW signals will persist in PTA measurements over the full duration of the observations.
In order to make robust astrophysical inferences of GW sources, we need to understand the effectiveness and limitations of our data analysis techniques.
Therefore, these techniques should be thoroughly tested on simulated data in which we understand the signal and noise processes that have been injected.
\par

The primary GW sources for PTAs are supermassive black hole binaries (SMBHBs) at the centers of merging galaxies \citep[e.g.][]{svc2008}.
Individually resolvable SMBMB sources will emit GWs at nearly-constant orbital frequencies in the PTA band \citep[see e.g.,][]{Jenet+2004}. As opposed to these ``continuous waves'' (CWs), an ensemble of unresolved sources will produce a stochastic gravitational wave background (GWB).
The expected signal for a first detection of GWs by pulsar timing is the GWB from SMBHBs \citep[e.g.][]{rsg2015}.
\par

The International Pulsar Timing Array \citep[IPTA;][]{IPTADR1} combines the effort of regional PTA collaborations to obtain the best sensitivity to low-frequency GWs by increasing the number of pulsars observed, extending the timing baselines for which they are observed, increasing the effective cadence of observations, and improving the angular coverage of our pulsars to build sensitivity over the whole sky \citep[e.g.][]{Siemens+2013,vs2016}.
One primary effort of the IPTA toward this goal is the development of new methods and strategies for GW data analysis \citep[e.g.][]{vH2013,Lentati+2013,vHv2014}.
In 2012, the IPTA released its first Mock Data Challenge (MDC), which resulted not only in a number of submissions, but also the development of several new methods now widely used in PTA data analysis.
Individual results of the first MDC are described in \cite{Cornish2012, ellis+2012, Taylor+2012, Taylor+2013, vH+2013}.
A summary of these efforts is given in \cite{IPTAMDC2}, which also defines the second IPTA mock data challenge (MDC2).
\par

This work describes a submission of analyses of IPTA MDC2 from a group based at West Virginia University in the United States.
This work was conducted under the time constraint of a submission deadline, and several of our results could be improved by rerunning them for more time or making small changes to the methods.
We choose to present our submission \textit{as is} and hope that the lessons learned can inform future GW analyses of PTA data.
This work is organized as follows:
in \S\ref{sec:mdc} we briefly describe MDC2;
we present our data analysis methods using the \enterprise software package in \S\ref{sec:analysis};
we describe our results on the open and closed datasets in \S\ref{sec:open} and \S\ref{sec:closed}, respectively;
finally, in \S\ref{sec:discuss} we discuss the implications of this work and future prospects.
\par

%% file: sec_mdc.tex
\input{tabs/mdc_table.tex}
\input{tabs/injected_values_table.tex}

IPTA MDC2 is fully described in \citet{IPTAMDC2}, but we summarize the relevant details here briefly.
Table~\ref{tab:mdc} describes the three open (group 1, a.k.a. \texttt{g1}) and three closed (group 2, a.k.a. \texttt{g2}) datasets provided in \citet{IPTAMDC2}.
Table~\ref{tab:MDC2injections} describes the injected parameter values for both the open and closed datasets.
The original release of the data only contained the injected parameters for the open datasets.
The injected parameters for the closed datasets were made available after the submission deadline to MDC2.
Each dataset contains simulated measurements of pulse arrival times for 33 pulsars observed over 15 years with an average 30 day cadence (185 times-of-arrival, or ``TOAs'', each).
The open datasets contain data at only one observing frequency while the closed datasets had two observing frequencies.
We chose not to analyze the \textit{a} datasets of group 1 which are evenly sampled, analyzing only the more realistic \textit{b} datasets which were not.
\par

For each pulsar, a given parameter file (\textit{par-file}) describes the model predicting pulse TOAs due to pulsar spin dynamics, astrometry, binary orbit (if applicable), clock corrections, and dispersion measure (integrated electron column density along line-of-sight to the pulsar) \citep{TEMPO2}.
Files containing a list of TOAs (\textit{tim-file}) describe the time-stamp and error at which the pulse arrived, the observatory, and the radio frequency at which the pulse was observed.
The TOAs were generated using \libstempo \citep{libstempo}.
The solar system ephemeris (SSE) used to generate the simulated TOAs was Jet Propulsion Laboratory's DE436 \citep{DE436}.
In addition to the parameters describing each pulsar, there was some known or unknown injection of a GW signal common to all pulsars as well as unmodeled noise in the TOAs requiring characterization. 
\par

The three open datasets comprising group 1 contain known GW signals.
Two of these datasets (\texttt{g1.d1} and \texttt{g1.d2}) contain injected GWBs taking the form of a power law spectrum that arises from many circular SMBHBs undergoing GW-driven evolution.
The amplitude of the injected GWBs differs between the two datasets.
Additionally, the noise properties of the datasets differ.
The third open dataset (\texttt{g1.d3}) contains the signal from a single resolvable SMBHB.
The injected source parameters for the open datasets were provided in a JSON database along with the data via GitHub\footnote{\url{https://github.com/ipta/mdc2/}}.
We present our findings of the signal searches in these datasets in \S\ref{sec:open}.
\par

The three closed datasets of group 2 contained unknown sources.
\citet{IPTAMDC2} informed MDC participants that one of the three contains only a GWB, another contains a GWB and CW, and a third contains two resolvable CWs.
Determining which sources were injected into which dataset was left for the analysts to determine.
Following the submission deadline, the signal parameters for group 2 were been posted as a JSON database in the appropriate GitHub repository.
For convenience we report the injected signals for both the open and closed datasets in table \ref{tab:MDC2injections}.
We present our findings for closed datasets in \S\ref{sec:closed}.

%% file: tabs/mdc_table.tex
\begin{table}[t]
    \scriptsize
    \caption{Dataset formats for IPTA MDC2. Each of the 33 pulsars were observed for 15 years with an average 30 day cadence. We use the acronyms WN = white noise, RN = red noise, GWB = gravitational wave background (SB = stochastic background in the original), CW = continuous wave (SS = single source in the original).
    In boldface we give our findings for the unknown signals in the group 2 closed datasets.
    \newline
    $^{*}$ We note that in \cite{IPTAMDC2} \texttt{g1.d3} was reported to have both a GWB and a single CW.}
    \label{tab:mdc}
    \begin{center}
    \begin{tabular}{@{} c|c|c|c @{}}
    \hline\hline
Group Dataset & Frequency (GHz) & Noise & Signals \\
\hline
\texttt{g1.d1} (b) & 1.44 & WN & GWB\\
\texttt{g1.d2} (b) & 1.44 & WN, RN & GWB\\
\texttt{g1.d3} (b) & 1.44 & WN & CW$^{*}$\\
\texttt{g2.d1} & 0.8, 1.44 & WN, RN & \textbf{GWB}\\
\texttt{g2.d2} & 0.8, 1.44 & WN, RN & \textbf{GWB+CW}\\
\texttt{g2.d3} & 0.8, 1.44 & WN, RN & \textbf{2$\times$CW}\\
            \hline
    \end{tabular}
    \end{center}
\end{table}

%% file: tabs/injected_values_table.tex
\begin{table*}[t]
    \scriptsize
    \caption{Injected values for each dataset in MDC2. Definitions for signal parameters are given in section \ref{sec:analysis.signal}. Here $A_{\mathrm{GWB}}$ is the amplitude of the GWB, $f_{\rm{GW}}$ is the GW frequency, $\mathcal{M}_c$ is the chirp mass of the binary, $h$ is the characteristic strain, $\theta$ and $\phi$ are the source position on the sky, $\iota$ is the orbital inclination of the binary, and $D$ is the luminosity distance of the source. The signal parameters for the closed data sets were taken from the JSON database that was placed in the appropriate GitHub repository following the close of the challenge.
    }
    \begin{center}
    \begin{tabular}{@{} ccccccccc @{}}
        \hline\hline
            Dataset & $A_{\mathrm{GWB}}$& $f_{\rm{GW}}$ & $\mathcal{M}_{c}$ & $h$ & $\phi_{\rm{GW}}$ & $\theta_{\rm{GW}}$ & $\iota$ & $D$\\
            & ($\times 10^{-15}$) & ($\times 10^{-9}$ Hz) & ($\times 10^{9}$ M$_{\odot}$) & ($\times 10^{-14}$) & (rad) & (rad) & (rad) & (Mpc)  \\
            \hline

            \texttt{g1.d1} (b) & 0.66 & -- & -- & -- & -- & -- & -- & -- \\
            \texttt{g1.d2} (b) & 1.3 & -- & -- & -- & -- & -- & -- & -- \\
            \texttt{g1.d3} (b) & -- & 6.1 & 5.0 & 4.46 & 4.07 & 1.43 & 0.44 & 65 \\
            \texttt{g2.d1} & 0.97 & -- & -- & -- & -- & -- & -- & -- \\
            \texttt{g2.d2} & 0.85 & 3.7 & 4.3 & 21.44 & 3.33 & 0.64 & 0.84 & 75.4 \\
            \texttt{g2.d3} & -- & 18 & 2 & 10.7 & 0.59 & 2.11 & 0.30 & 121 \\
            & -- & 2.4 & 0.91 & 1.07 & 4.28 & 1.40 & 1.51 & 85 \\
            \hline
    \end{tabular}
    \end{center}
    \label{tab:MDC2injections}
\end{table*}

%% file: sec_analysis.tex
In general, we followed the methods used by the North American Nanohertz Observatory for Gravitational Waves (NANOGrav) to analyze their 11-year data set described in \citet{ng11.gwb} and \citet{ng11.cw}, as well as additional CW methods developed in \citet{wittinprep}.
We used NANOGrav's flagship data analysis software \enterprise \citep{enterprise} to calculate the likelihood and prior in our parameter space.
We used \ptmcmc \citep{ptmcmc} to perform Markov chain Monte Carlo (MCMC) sampling of this posterior distribution.
We used these samples to perform Bayesian parameter estimation and inference.
\par

While \ptmcmc is capable of performing Parallel Tempered MCMC, we did not use this capability.
\ptmcmc provides several built-in proposal distributions, including Adaptive Metropolis, Single Component Adaptive Metropolis, and Differential Evolution methods\footnote{See the \ptmcmc documentation for more details: \url{http://jellis18.github.io/PTMCMCSampler/}.}.
We used a blend of these built-in proposal schemes and occasionally proposed moves drawn from the search parameter prior distributions to aid mixing.
Additionally, we provided \ptmcmc with a list of ``sampling groups'' which collect parameters with known correlations.
\ptmcmc can then propose simultaneous moves for grouped parameters to increase acceptance.
\par

In the \enterprise framework we defined parameterized \textit{signals} that model the noise and GWs in the data.
\enterprise models stochastic signals as Gaussian processes\footnote{See the \enterprise documentation for more details: \url{https://enterprise.readthedocs.io/en/latest/}.}.
We describe the models we used below and summarize the \enterprise signals in Table \ref{tab:sig}.

\input{tabs/signals_table.tex}

\subsection{Noise Model}
\label{sec:analysis.noise}

For each pulsar we used the same basic noise model containing both white noise (WN) and red noise (RN).
The noise covariance matrix is written as
\be 
    \mathbf{C}_{t t^\prime} =
        \delta_{t t^\prime}
        \left(\mathcal{F}^2\sigma_{\mathrm{S/N}}^2 + \mathcal{Q}^2\right) 
        + \mathbf{C}_{\red,t t^\prime}.
    \label{eq:noise}
\ee
WN originates from template fitting error, $\sigma_\mathrm{S/N}$, which is related to the signal-to-noise ratio of an average pulse profile.
An additional factor $\mathcal{F}$ multiplies the base template fitting error (EFAC).
Excess WN is added in quadrature as $\mathcal{Q}$ (EQUAD)\footnote{Note that in the literature, depending on the timing software used, EFAC may multiply the quadrature sum including the EQUAD and not only the template-fitting errors.}.
The Kronecker delta function is given by $\delta_{t t^\prime}$.
\par

RN in individual pulsars is described by a power law spectrum with amplitude $A_\red$ (in units of $\mu\mathrm{s~yr}^{1/2}$), spectral index $\gamma_\red$, and referenced to a frequency of $1~\yr^{-1}$, written as
\be 
    S_\red(f) = A_\red^2 \left(\frac{f}{\yr^{-1}}\right)^{\gamma_\red}.
\ee
The equivalent covariance matrix is of the form
\be 
    \mathbf{C}_{\red,t t^\prime} = \int_{1/T}^\infty S_{\red}(f)\cos(2\pi f t)~\dd f.
    \label{eq:red}
\ee

The complete noise model had four search parameters per pulsar $(\mathcal{F},\mathcal{Q}, A_\red, \gamma_\red)$.
We expected the RN to be covariant with any GW signal, so we always simultaneously fit RN and GW models.
The WN should not be covariant with GWs, so there are two ways to handle those parameters.
First we could let them be search parameters that are fit simultaneously with the per pulsar RN and GWs.
In this case the search space can be of very high dimension.
The second option is to prefit the WN parameters by analyzing individual pulsars separately, then fix the WN parameters to their best fit (median) value for the GW analysis.
For CW searches, which already have a large number of parameters (see \S\ref{sec:analysis.signal.cw}), the second method is preferred.

\subsection{GW Model}
\label{sec:analysis.signal}

\subsubsection{Gravitational Wave Background}
\label{sec:analysis.signal.gwb}
Several models for a stochastic GWB have been considered in the literature, but we used the most basic form of a power law due to SMBHBs in circular orbits that are inspiraling due to GW emission only.
In this case the GW cross-power spectral density between two pulsars, $a$ and $b$, is
\be
    S(f)_{ab} = \Gamma_{ab} \frac{A_\GWB^2}{12\pi^2}
        \left(\frac{f}{f_\yr}\right)^{-\gamma} \,\yr^3,
    \label{eq:gwb}
\ee
where $\Gamma_{ab}$ is the overlap reduction function describing the angular correlation pattern between pulsars, $A_\GWB$ is the characteristic strain amplitude, and $\gamma$ is the spectral index of the timing residuals.
Assuming purely GW driven evolution, the GWB spectral index is $\gamma = 13/3$.
In general we could search over the spectral index; however, we chose to hold $\gamma$ fixed to the expected value for all GWB analyses.
\par

In our analyses we compared two overlap reduction functions for the GWB:
\begin{itemize}[noitemsep]
    \item Hellings--Downs correlations (HD), which describes quadrupolar spatial correlations due to GWs \citep{hd1983};
    \item Uncorrelated common red noise (CRN) process, where $\Gamma_{ab} = \delta_{ab}$.
\end{itemize}
A true GWB should be determined by the telltale HD signature.
If we determine a GWB detection by comparing an HD GWB model to fully individualized per pulsar noise however, other common noise processes (e.g. clock errors, SSE errors) could be mistaken for HD correlations \citep{Tiburzi+2016}.
The gold standard of detection should compare the significance of an HD GWB against a generic CRN process to prove the existence of spatial correlations.
\par

The GWB model appears in the noise covariance matrix as an additional source of red noise in the same way as Eq. \ref{eq:red}.
In the case of the CRN overlap reduction function the total covariance matrix is block diagonal, with one block per pulsar.
For the HD overlap reduction function, the covariance has off-diagonal terms owing to inter-pulsar correlations.
\par

Because we chose to hold the spectral index of the GWB fixed at $\gamma = 13/3$ the GWB model has only one search parameter: the strain amplitude of the spectrum $A_\GWB$ at $f=1~\yr^{-1}$.

\subsubsection{Continuous Waves}
\label{sec:analysis.signal.cw}

CW sources produce deterministic signals, thus we can perform an effective matched filter search.
We described CW sources using eight parameters, as listed in Table \ref{tab:MDC2injections} \citep{sv2010}: 
\begin{itemize}[noitemsep]
    \item source position on the sky $(\theta, \phi)$;
    \item GW frequency, related to the orbital frequency of the binary at some reference time $(f)$;
    \item orbital phase of the binary at some reference time $(\varphi_0)$;
    \item orbital inclination of the binary $(\iota)$;
    \item chirp mass of the binary $(\mathcal{M}_c)$;
    \item GW polarization angle $(\psi)$;
    \item characteristic strain, which is related to the chirp mass, GW frequency, and luminosity distance of the source $(h)$
\end{itemize}
These parameters describe the CW signal at the Earth, often called the \textit{Earth term}.
We also included the \textit{pulsar term} in our search, which depends on the frequency and phase of the CW at each pulsar.
The pulsar term is determined by the system chirp mass, reference frequency, and the source-pulsar geometry including sky positions and pulsar distance.
For the pulsar term we used the: 
\begin{itemize}[noitemsep]
    \item sky position of each pulsar, whose uncertainties are small enough to be considered exact;
    \item distance to each pulsar ($d_\mathrm{psr}$);
    \item CW phase at each pulsar ($\varphi_\mathrm{psr}$).
\end{itemize}
The pulsar distances are not well measured \textit{a priori}, so they are left as search parameters.
The CW phase at the pulsar is determined by the source-pulsar geometry, so the CW phase is not strictly needed.
Unfortunately, the likelihood surface for the pulsar distance is highly complex, causing practical sampling issues.
Following \citet{ellis+2013} and \citet{ng11.cw}, we used the pulsar distance to determine the GW frequency at the pulsar, but fit the phase independently.
For a more thorough description, see \citet{ng11.cw} and references therein. 

The CW model has eight global parameters, plus two additional parameters per pulsar.
For the 33-pulsar MDC2 data, a CW search requires 74 parameters.
This is in addition to any noise or GWB parameters that must also be fit.
\par

\subsection{Handling uncertainty in pulsar and SSE ephemerides}
\label{sec:analysis.other}

Following the scheme of \citet{vH2013}, \enterprise can account for the uncertainty in the pulsar timing model parameters found in each \textit{par-file} without adding additional search parameters.
This is done by using a linearized, quadratic timing model and analytically marginalizing over the timing uncertainties before computing the likelihood.
\par

A stochastic gravitational wave background is covariant with uncertainties in the SSE \citep{Tiburzi+2016, ng11.gwb, ipta.sse}
Since the MDC data were produced with a known SSE (JPL's DE436), we used this SSE in our analyses of closed datasets.
In our analysis of open datasets, \S\ref{sec:open.g1.d12}, we considered how SSE uncertainty alters detection prospects by implementing the \bayesephem model \citep{ng11.gwb}.
\bayesephem perturbs a given SSE in a deterministic way, changing the location of the solar system barycenter.
Perturbations to the masses of the gas and ice giants and Jupiter's Keplerian orbital elements are parameters in the model.
\par

\subsection{Bayesian Inference}

For all analyses we report a Bayes factor, $\mathcal{B}$, as a measure of detection significance.
All Bayes factors were estimated using the Savage-Dickey method which compares the prior probability to the posterior probability for low GW amplitudes \citep{Dickey1971}.
We place a uniform prior on the log (base-10) amplitude of GW signals, i.e., $\log A_\GWB$ and $\log h_c$.
To estimate the the posterior probability for low amplitude, we used samples from the few lowest-amplitude bins.
\par

In cases where there are no samples in the low amplitude bins, the Bayes factor cannot be estimated with this method.
Instead we report a value of $\mathcal{B} = \infty$ to represent a lower limit $\mathcal{B} > \mathrm{Pr}\cdot N_\mathrm{sample}$, where $N_{\mathrm{sample}}$ is the number of MCMC samples in an individual analysis and $\mathrm{Pr}$ is the uniform prior probability.
\par

We declare a ``detection" for Bayes factors $>3$.
This corresponds to 3:1 odds or 75\% confidence, which is quite a low threshold.
To declare a detection in practice we would hope for a Bayes factor $>100$.
For GWB detections, we report the measured GWB amplitude as both the median with a 90\% credible interval and the mode of the posterior distribution.
For strong detections, the median and mode should agree.
For marginal detections where the posterior has long tails, the mode is a better representation of the ``best fit'' as it is the maximum posterior value.
In cases where we did not detect a GWB we present 95\% upper limits on the GWB amplitude.
\par

For CW analyses we report measured parameters as the median with a 90\% credible interval.
For non-detections we present an upper limit on the CW amplitude and do not report other parameters.
\par

We used log-uniform priors on GWB and CW amplitudes throughout our analyses.
This choice of prior is suitable for estimating Bayes factors by the Savage-Dicke approximation.
However, we cannot integrate our posteriors all the way down to $A=0$ ($\log A\rightarrow -\infty$).
Because of this any upper limits we present depend on our choice of prior range.
We used consistent priors throughout, but our upper limits are not strictly speaking ``robust'' \citep{ng11.gwb}.

%% file: tabs/signals_table.tex
\begin{table*}[t]
    \scriptsize
    \caption{\enterprise signals and their parameters.
        Not all analyses use all of these signals.
        See the results for particular datasets for details.
        Signal type refers to the base class of the signal in the \texttt{enterprise.signals} submodule.}
    \label{tab:sig}
    \begin{center}
    \begin{tabular}{@{} c|c|c|c @{}}

Signal Type & Parameter Name & Per-Pulsar Parameters &  Global Parameters \\
\hline\hline
\multirow{2}{*}{\texttt{WhiteNoise}} & EFAC & $\mathcal{F}$ & -- \\
 & EQUAD & $\mathcal{Q}$ & -- \\
\hline
\texttt{FourierBasisGP} & Pulsar Red Noise & $A_\red, \gamma_\red$ & -- \\
\hline
\texttt{FourierBasisCommonGP} & GWB & -- & $A_\GWB, \gamma=13/3$ \\
\hline
\multirow{2}{*}{\texttt{Deterministic}} & CW & $d_\mathrm{psr}, \varphi_\mathrm{psr}$ & $h, f, \mathcal{M}_c, \iota, \theta, \phi, \psi, \varphi_0$ \\
 & \bayesephem & -- & $M_\mathrm{Jup}, M_\mathrm{Sat}, M_\mathrm{Ur}, M_\mathrm{Nep}, \alpha_{\mathrm{Jup}, i}$ \\
\hline
\texttt{BasisGP} & Timing Model & -- & -- \\
\hline\hline

    \end{tabular}
    \end{center}
\end{table*}

%% file: sec_open.tex
In order to verify our methods, we analyzed the unevenly sampled group 1 \textit{b} datasets.
We break analyses and discussion into two parts based on the presence of a CW signal.
\par

\subsection{Open datasets \texttt{g1.d1}-GWB+WN and \texttt{g1.d2}-GWB+WN+RN}
\label{sec:open.g1.d12}

Both datasets \texttt{g1.d1} and \texttt{g1.d2} contain only a stochastic GWB signal.
The key difference between these two datasets is the presence of per-pulsar RN in \texttt{g1.d2}, while \texttt{g1.d1} has purely WN.
For each dataset we used a base model consisting of marginalized timing uncertainty, WN, RN, and an HD correlated GWB.
All of the model parameters were fit simultaneously, giving $33\times4 + 1 = 133$ free parameters.
To verify our GWB analysis methods we performed several analyses on both of these datasets.

\begin{figure}
    \centering
    \includegraphics[width=\linewidth]{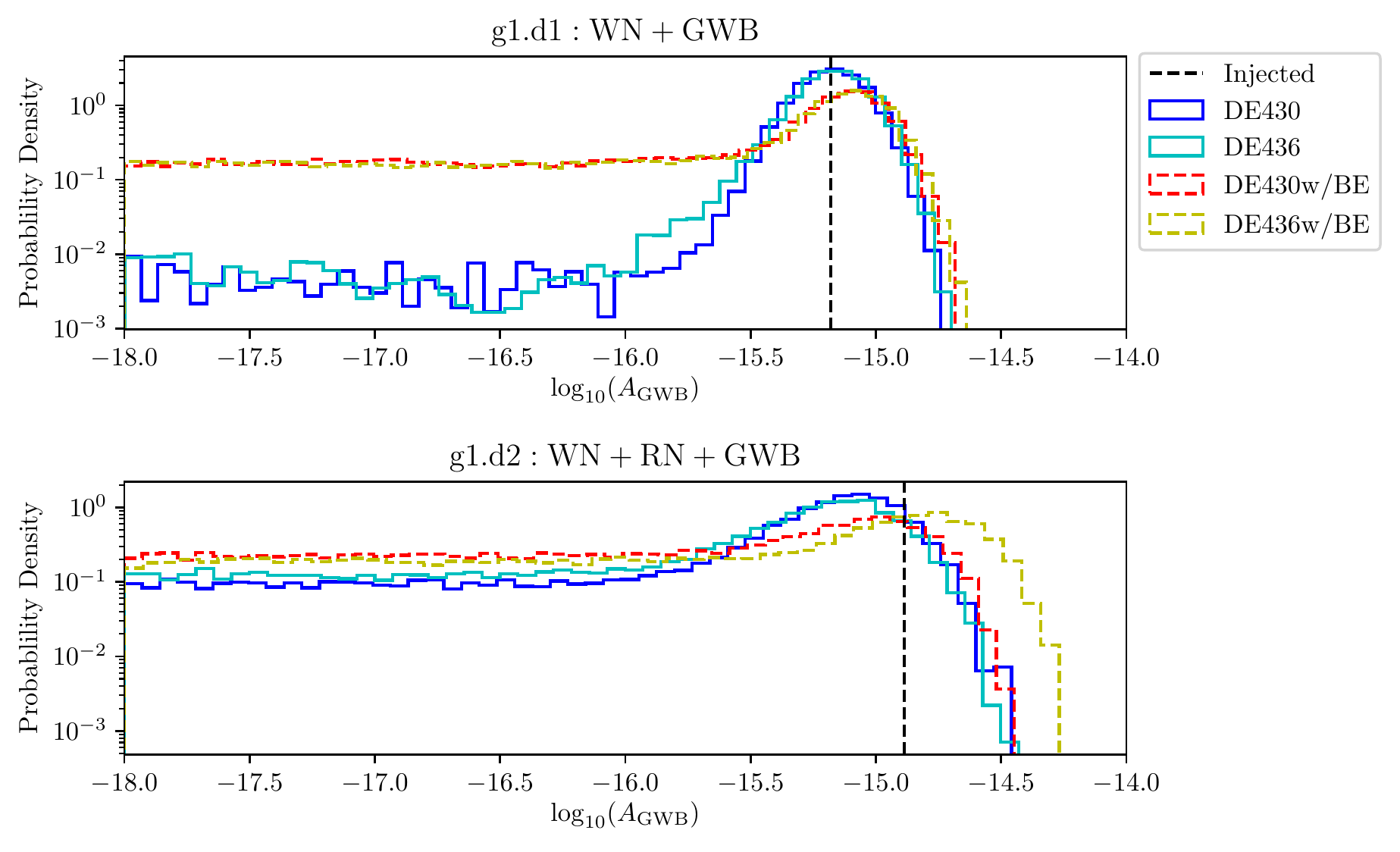}

    \caption{GWB amplitude posterior with Hellings--Downs spatial correlations comparing different SSEs with and without \bayesephem (BE) for open datasets \texttt{g1.d1} and \texttt{g1.d2}.
    }
    \label{fig:g1.d12eph}
\end{figure}

To test the performance of \bayesephem, we analyzed these two datasets using the base model described above with two different SSEs:
first the input one, DE436; second DE430 \citep{DE430}, an older JPL ephemeris.
This allowed us to determine how SSE errors affected the significance of a GWB detection.
We also repeated the analyses using \bayesephem in addition to the base model for both SSEs.
The posterior distribution of the GWB amplitude for these analyses are shown in Figure~\ref{fig:g1.d12eph}.
The results of these analyses are reported in Tables \ref{tab:open1} and \ref{tab:open2}.

For all datasets and analyses the mode of the posterior distribution fell near the injected amplitude.
However, the significance of the detections varied widely.
For \texttt{g1.d1} we achieved a strong detection only when using a fixed SSE with no \bayesephem perturbations. The fact that the inclusion of \bayesephem weakens the detection significance is not unexpected; the signature induced in pulsar timing residuals by an inaccurate solar system ephemeris is similar to that induced by a GWB and so some fraction of the injected GWB is absorbed by \bayesephem.
For either choice of SSE, using \bayesephem yielded very similar posteriors on $A_\GWB$, but the GWB was not detected.
What is surprising, is that using the incorrect SSE we still found a strong detection of the background.

\par

For \texttt{g1.d2} we found no significant GWB under any search method.
Again, \bayesephem led to nearly identical posteriors regardless of initial SSE.
The largest Bayes factor came from the correct choice of SSE and the weakest from the incorrect.
These results from \texttt{g1.d2} match our intuition better than those from \texttt{g1.d1}.
We report upper limits on the GWB amplitude for these runs in Table \ref{tab:open2}.
The `best fit' posterior modes should be taken with a grain of salt, given the associated Bayes factors.

During the SSE comparisons for datasets \texttt{g1.d1} and \texttt{g1.d2}, we noticed some irregularities in the posteriors for noise parameters.
In a small number pulsars the use of a different SSE or the addition of \bayesephem leads to vastly different noise parameter posteriors, where the posterior distribution is constrained to a significantly smaller fraction of the parameter space.
In \ref{sec:appendix.eph} we discuss some of these issues in greater detail.
\par

\begin{figure}
    \centering
    \includegraphics[width = \linewidth]{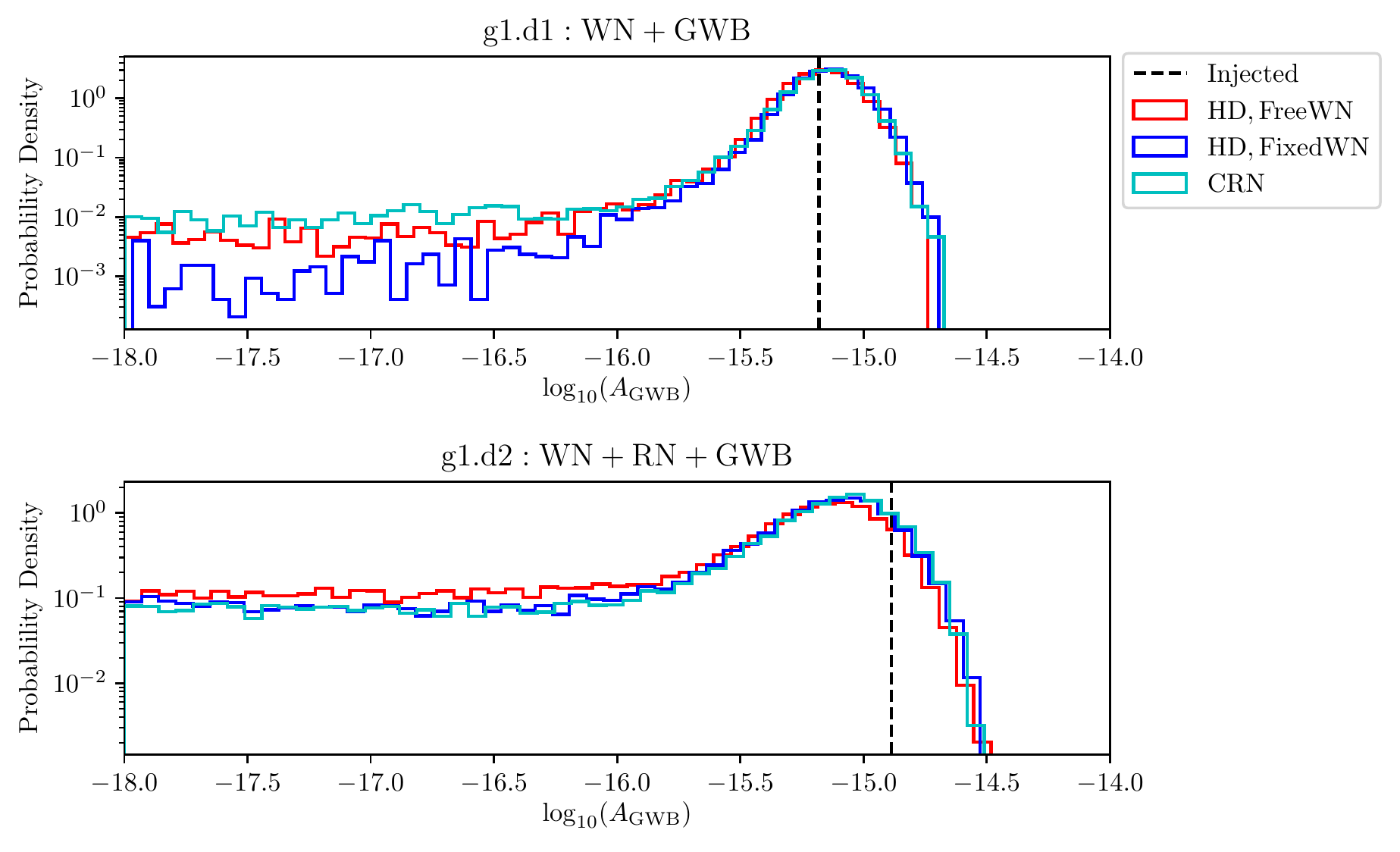}
    \caption{Gravitational wave background (GWB) amplitude distributions comparing a detection for correlated red noise (CRN) with two distinct Hellings--Downs correlations (HD): one with fixed pulsar white noise (WN) parameters and the other with simultaneously searched WN parameters.
    The injected GWB amplitude is shown with the dashed black line. 
    }
    \label{fig:g1.d12_wn}
\end{figure}

To investigate how prefitting for WN parameters on a per-pulsar basis versus sampling over them during the GWB detection run affects the GWB search, we implemented two additional models to search \texttt{g1.d1} and \texttt{g1.d2} for the stochastic GWB.
Both models consisted of the marginalized timing uncertainty, WN, RN, and HD correlations as described above.
The first model used uniform \enterprise priors on WN parameters as in \citet{ng11.gwb} (`free WN'), while the second model set those parameters to constant values determined by individual pulsar noise analyses (`fixed WN').
These prefitting noise analyses largely recovered the injected WN parameters.
In \texttt{g1.d1}, 65\% of the recovered WN parameter values agreed with the injected values to within 5\% and all but one agreed to within 15\%, the only discrepancy being the recovered EFAC for PSR~J1022+1001, which was overestimated by 25\%.
In \texttt{g1.d2}, 64\% of the recovered values agreed with the injected values to within 5\% and all but two agreed to within 15\%.
The recovered EFAC and EQUAD for PSR~J1024--0719 were greatly overestimated, colliding with the upper edge of their prior ranges.
This is very similar to unusual noise recovery discussed in \ref{sec:appendix.eph}.
As can be seen in Figure \ref{fig:g1.d12_wn} and Tables \ref{tab:open1} and \ref{tab:open2}, these two methods for handling WN agree with each other with respect to GWB recovery.
This is expected, as WN should not be covariant with the red GWB.
This result further validates the standard NANOGrav practice, where WN parameters are prefit to reduce the dimensionality of GW analyses.

\input{tabs/open1_table.tex}
\input{tabs/open2_table.tex}

\subsection{Open dataset \texttt{g1.d3}-CW+WN}
\label{sec:open.g1.d3}

In \cite{IPTAMDC2} it was stated that dataset \texttt{g1.d3} contained both a GWB and a CW and as such we performed three different analyses of dataset \texttt{g1.d3}, all of which searched for both the stochastic GWB and a single-source CW signal.
After the close of MDC2, it was determined that dataset \texttt{g1.d3} did not in fact contain a GWB, and therefore our searches in this dataset may not have been optimal for recoving the injected signals.
However the methods and results presented here still yield interesting results worthy of further discussion.
All three analyses included the marginalized timing uncertainty, WN, GWB, and CW.
In all cases, we searched for the GWB using Hellings--Downs spatial correlations and the DE436 ephemeris without \bayesephem.
Note that even though there was no red noise injected into this dataset, we included it in our initial pulsar noise runs to account for the GWB.
The three analyses differed in their treatment of WN. The three methods were:
\begin{enumerate}
    \item[(1)] including the WN parameters as free parameters in the GW search;
    \item[(2)] prefitting the WN parameters based on individual pulsar analyses (fixed w/o CW);
    \item[(3)] prefitting the WN parameters based on individual pulsar analyses that included a CW signal (fixed w/ CW).
\end{enumerate}
Method 1 used 141 search parameters, while Methods 2 and 3 had only 75 due to the fixed WN.
\par

Results for the three CW + GWB methods are shown in Tables \ref{tab:open3} and \ref{tab:open3CW}.
Additionally, we performed GWB-only analyses following the methods of \S\ref{sec:open.g1.d12}, and we report these results in Table \ref{tab:open3}.
Corner plots showing the 1D and 2D posteriors for the individual CW parameters can be found in \ref{sec:appendix.cw}.
\par

\input{tabs/open3_GWB_table.tex}
\input{tabs/open3_CW_table.tex}

Methods 1 and 3 most accurately recovered the CW parameters and the results for these two methods are consistent as shown in \autoref{tab:open3CW}. However, Method 1 was significantly more computationally expensive owing to the addition of the two WN parameters per pulsar.
This method is likely not practically feasible for use on real datasets in the future.
See \S\ref{sec:discuss} for more discussion on this point.
\par

Method 2 recovered the CW parameters less accurately, showing the need to include the CW model in prefitting noise analyses.
The CW signal is loud enough that it biases the WN parameters when it is not included.
Strangely, this method did detect a GWB despite the fact that no GWB was injected in the dataset\footnote{We note again that in the initial presentation of MDC2 a GWB was reported to have been injected in this dataset.}.
We suspect the GWB model is fitting leaked power from the CW and noise models, because the WN parameters had been fixed to incorrect values.

Our detection of a GWB using Method 2 highlights an important failure mode of that method.
The noise prefitting assumes that the data is well described by a noise only model.
In the case when there is a detectable CW signal present, this assumption is violated.
The mismatch of the model and data leads to the unexpected consequence of detecting a GWB signal that was not there.
\par

A recent search for CWs, \citet{ng11.cw}, followed the procedure of method 2, but did not detect anything.
In their case, the assumption that the data is well described by the noise model alone was not violated.
Our mistaken detection can serve as a cautionary tale:
in the future the detection of a CW source should only be trusted if it can be verified using methods akin to our Methods 1 or 3.
\par

%% file: tabs/open1_table.tex
\begin{table}[t]
    \scriptsize
    \caption{Detection comparison for \texttt{g1.d1}.
            Bayes factors are for GW and noise vs. noise only,
            and the uncertainties on median are 90\% CIs.
            Ephemeris comparison runs with \bayesephem are denoted ``w/ BE''.
            We report the result of modeling common red noise with no spatial correlations (CRN) and with spatial correlations (HD). The injected amplitude was $0.7 \times 10^{-15}$.}
    \begin{center}
    \begin{tabular}{@{} cccc @{}}
        \hline\hline
            run type & $A_{\rm{med}}\; (\times10^{-15})$ & $A_{\rm{mode}}\; (\times10^{-15})$ & $\mathcal{B}$ \\
            \hline
            \multicolumn{4}{c}{Common Red Noise Analysis (DE436, free WN)}\\
            \hline
            CRN & $0.7^{+0.4}_{-0.4}$ & $0.8$  & $23$ \\
            \hline
            \multicolumn{4}{c}{HD with White-Noise Comparison (DE436)}\\
            \hline
            HD (fixed WN) & $0.7^{+0.4}_{-0.3}$ & $0.8$  & $\infty$ \\
            HD (free WN) & $0.7^{+0.4}_{-0.3}$ & $0.7$  & $40$ \\
            \hline
            \multicolumn{4}{c}{HD with Ephemeris Comparison (free WN)}\\
            \hline
            DE436       & $0.7^{+0.4}_{-0.3}$ & $0.6$  & $27$ \\
            DE436 w/ BE & $<1.2$ & $0.8$  & $1.2$ \\
            DE430 w/ BE & $<1.2$ & $0.8$  & $1.2$ \\
            DE430       & $0.7^{+0.4}_{-0.3}$ & $0.7$  & $39$ \\
            \hline\hline
    \end{tabular}
    \end{center}
    \label{tab:open1}
\end{table}

%% file: tabs/open2_table.tex
\begin{table}[t]
    \scriptsize
    \caption{Detection comparison for \texttt{g1.d2}. See Table~\ref{tab:open1} caption for more information. The injected amplitude was $1.3 \times 10^{-15}$. Numerical differences between the HD with free WN row and DE436 row (rows 3 and 4) result from two separate runs reported.}
    \begin{center}
    \begin{tabular}{@{} cccc @{}}
        \hline\hline
            run type & $A_{\rm{med}}\; (\times10^{-15})$ & $A_{\rm{mode}}\; (\times10^{-15})$ & $\mathcal{B}$ \\
            \hline
            \multicolumn{4}{c}{Common Red Noise Analysis (DE436, free WN)}\\
            \hline
            CRN & $<1.5$ & $0.9$  & $2.6$ \\
            \hline
            \multicolumn{4}{c}{HD with White-Noise Comparison (DE436)}\\
            \hline
            HD (fixed WN) & $<1.4$ & $0.9$  & $2.3$ \\
            HD (free WN) & $<1.3$ & $0.8$  & $1.8$ \\
            \hline
            \multicolumn{4}{c}{HD with Ephemeris Comparison (free WN)}\\
            \hline
            DE436       & $<1.4$ & $0.9$  & $1.6$ \\
            DE436 w/ BE & $<2.3$ & $1.8$  & $1.1$ \\
            DE430 w/ BE & $<1.4$ & $1.0$  & $0.9$ \\
            DE430       & $<1.4$ & $0.9$  & $2.1$ \\
            \hline
    \end{tabular}
    \end{center}
    \label{tab:open2}
\end{table}

%% file: tabs/open3_GWB_table.tex
\begin{table}[t]
    \scriptsize
    \caption{Detection comparison for \texttt{g1.d3}.
            We report the GWB amplitude, mode, and Bayes factor recovered for all three different searches including a CW, as a GWB was search for in each method, though it was later determined that a GWB was not injected in this dataset.
            Bayes factors are for GW and noise vs. noise only,
            and the uncertainties on median are 90\% CIs.
            We report the results of modeling common red noise with no spatial correlations (CRN) and with spatial correlations (HD) without a CW included.
            }
    \begin{center}
    \begin{tabular}{@{} cccc @{}}
        \hline\hline
            run type & $A_{\rm{med}}\; (\times10^{-15})$ & $A_{\rm{mode}}\; (\times10^{-15})$ & $\mathcal{B}$ \\
            \hline
            CRN & $<3.3$ & $3.2$  & $0.7$ \\
            HD & $<4.7$ & $3.4$  & $1.5$ \\
            \hline
            HD + CW (1) & $ < 0.28$ & $0.06$ & $0.5$ \\
            HD + CW (2) & $6^{+2}_{-2}$ & $5.66$ & $\infty$ \\
            HD + CW (3) & $ < 0.27$ & $0.00$ & $0.4$ \\
            \hline
    \end{tabular}
    \end{center}
    \label{tab:open3}
\end{table}

%% file: tabs/open3_CW_table.tex
\begin{table*}[t]
    \scriptsize
    \caption{Continuous gravitational wave detection comparison for \texttt{g1.d3}.
    }
    \begin{center}
    \begin{tabular}{@{} ccccccccc @{}}
        \hline\hline
            run type & $f_{\rm{GW}}$ & $\mathcal{M}_{c}$ & $h$ & $\phi_{\rm{GW}}$ & $\theta_{\rm{GW}}$ & $\iota$ & $D$ & $\mathcal{B}$\\
            & ($\times 10^{-9}$ Hz) & ($\times 10^{9}$ M$_{\odot}$) & ($\times 10^{-14}$) & (rad) & (rad) & (rad) & (Mpc) &  \\
            \hline
            Injected Values & 6.1 & 5.0 & 4.46 & 4.07 & 1.43 & 0.44 & 65 & -- \\
            \hline
            1 (free WN) & $6.07^{+0.05}_{-0.05}$ & $4.5^{+0.6}_{-0.6}$ & $4.8^{+1.4}_{-0.8}$ & $4.07^{+0.04}_{-0.04}$ & $1^{+1}_{-1}$ & $0.6^{+0.3}_{-0.4}$ & $60^{+20}_{-20}$ & $\infty$ \\
            2 (fixed WN w/o CW) & $6.02^{+0.1}_{-0.1}$ & $2^{+1}_{-2}$ & $8^{+3}_{-3}$ & $4.98^{+0.03}_{-0.06}$ & $2^{+2}_{-2}$ & $2.1^{+0.6}_{-0.2}$ & $10^{+20}_{-10}$ & $\infty$  \\
            3 (fixed WN w/ CW) & $6.07^{+0.05}_{-0.06}$ & $4.6^{+0.7}_{-0.8}$ & $4.7^{+1.2}_{-0.7}$ & $4.07^{+0.04}_{-0.04}$ & $1^{+1}_{-1}$ & $0.6^{+0.3}_{-0.4}$ & $70^{+20}_{-20}$ & $\infty$ \\
            \hline
    \end{tabular}
    \end{center}
    \label{tab:open3CW}
\end{table*}

%% file: sec_closed.tex
The closed datasets were intended to act as a more realistic test of signal recovery for the methods we verified with the open datasets. 
All three of the closed datasets contained simulated pulsar white noise and red noise. 
In \citet{IPTAMDC2}, the types of GW signals used for the closed sets were listed but the specific datasets to which they corresponded were not.
Within the closed datasets there is a dataset with a single stochastic GWB, one with a stochastic GWB and a single CW, and one with two CWs and no GWB.
\par

For each dataset we ran a series of three analyses to determine which contained which GW signals.
We compared two GWB models (defined in \S\ref{sec:analysis.signal.gwb}): uncorrelated, common red noise (CRN) and Hellings--Downs, quadrupolar correlated red noise (HD).
Both of these models included directly sampled, i.e. `free',  WN parameters.
We also ran a GWB + CW analysis following method 3.
After reviewing the results of these analyses we were able to determine which datasets contained which sources, and our findings are stated in Table \ref{tab:mdc}.
We did not perform a formal model section step.

\begin{figure}
    \centering
    \includegraphics[width = \linewidth]{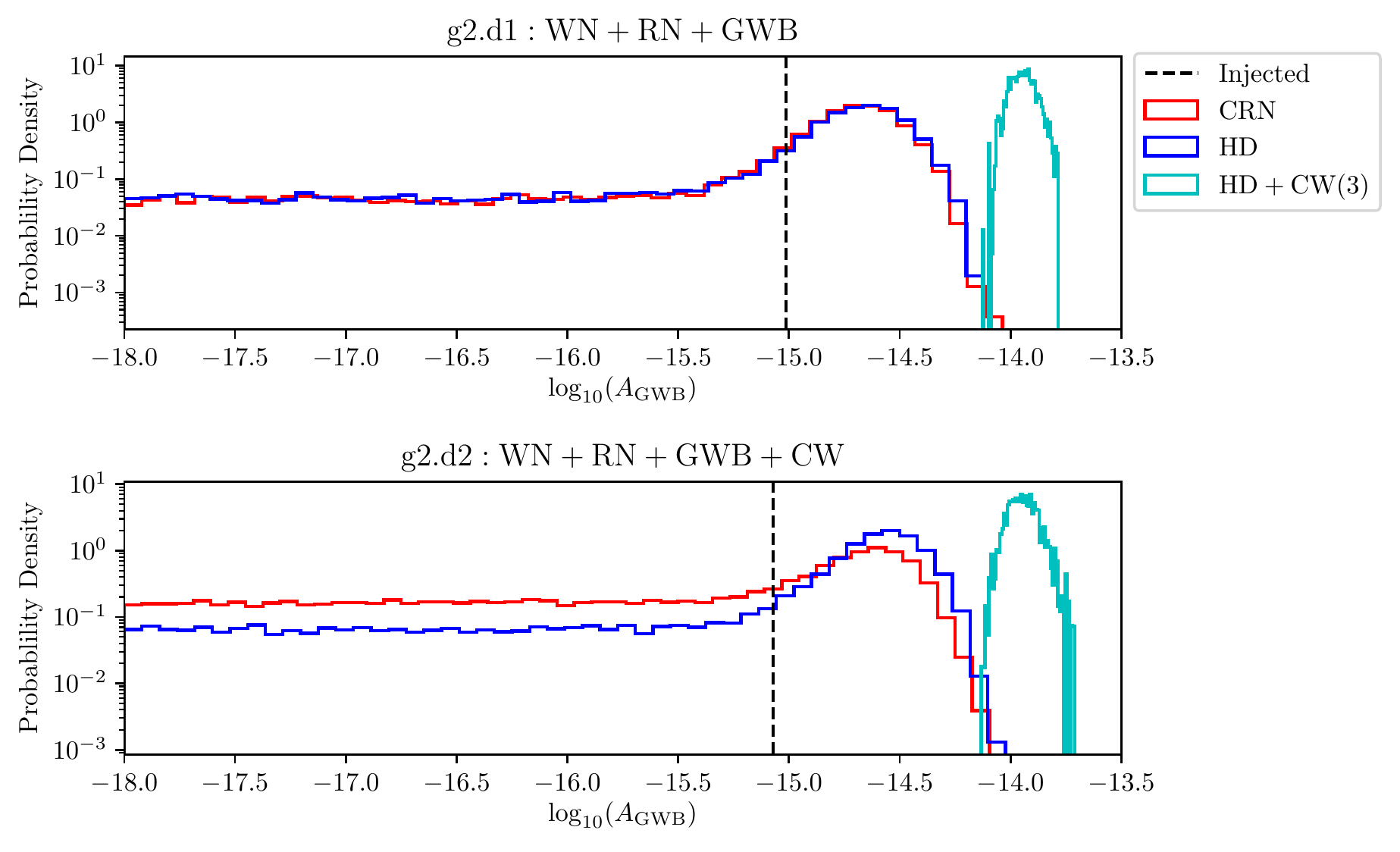}

    \caption{GWB amplitude posterior for closed dataset analyses.
    For \texttt{g2.d2} we include the GWB posterior from a method 1 GWB + CW analysis (see \S\ref{sec:open.g1.d3}), in addition to searches for GWB only.
    }
    \label{fig:g2.d12}
\end{figure}

\subsection{Closed Dataset \texttt{g2.d1}}
We found agreement between posterior distributions of the GWB amplitude for the analyses using either a CRN or HD model (see Table \ref{tab:closed1}).
However, our detection of the GWB was marginal.
In addition to GWB only searches, we ran a method 3 type GWB + CW search (see \S\ref{sec:open.g1.d3}) on this closed dataset, which showed strong evidence for a high amplitude GWB but no CW.
We suspect that the unsupported CW model introduced residual power for which the GWB model was forced to compensate.
This is another case where the model being mismatched to the data leads to incorrect findings.
We concluded that this dataset contains solely a GWB.

\input{tabs/closed1_table.tex}

\subsection{Closed Dataset \texttt{g2.d2}}
\label{sec:closed.g2.d2}

Using the same detection methods as the first closed dataset, we found very weak evidence for a GWB (see Table \ref{tab:closed2GWB}). 
In the case of the method 3 type GWB + CW search, we did detect a CW signal.
The parameters for the CW search are shown in Table \ref{tab:closed2CW}.
This analysis was unable to converge upon a chirp mass, returning a nearly uniform posterior covering the entire prior range from $10^7 - 10^{10} ~ \rm{M}_\odot$.
This is indicative of our search only detecting the Earth term.
The GW frequency at the pulsar, which is encoded in the pulsar term, is needed to constrain the chirp mass.
\par

\begin{figure}
    \centering
    \includegraphics[width = \textwidth]{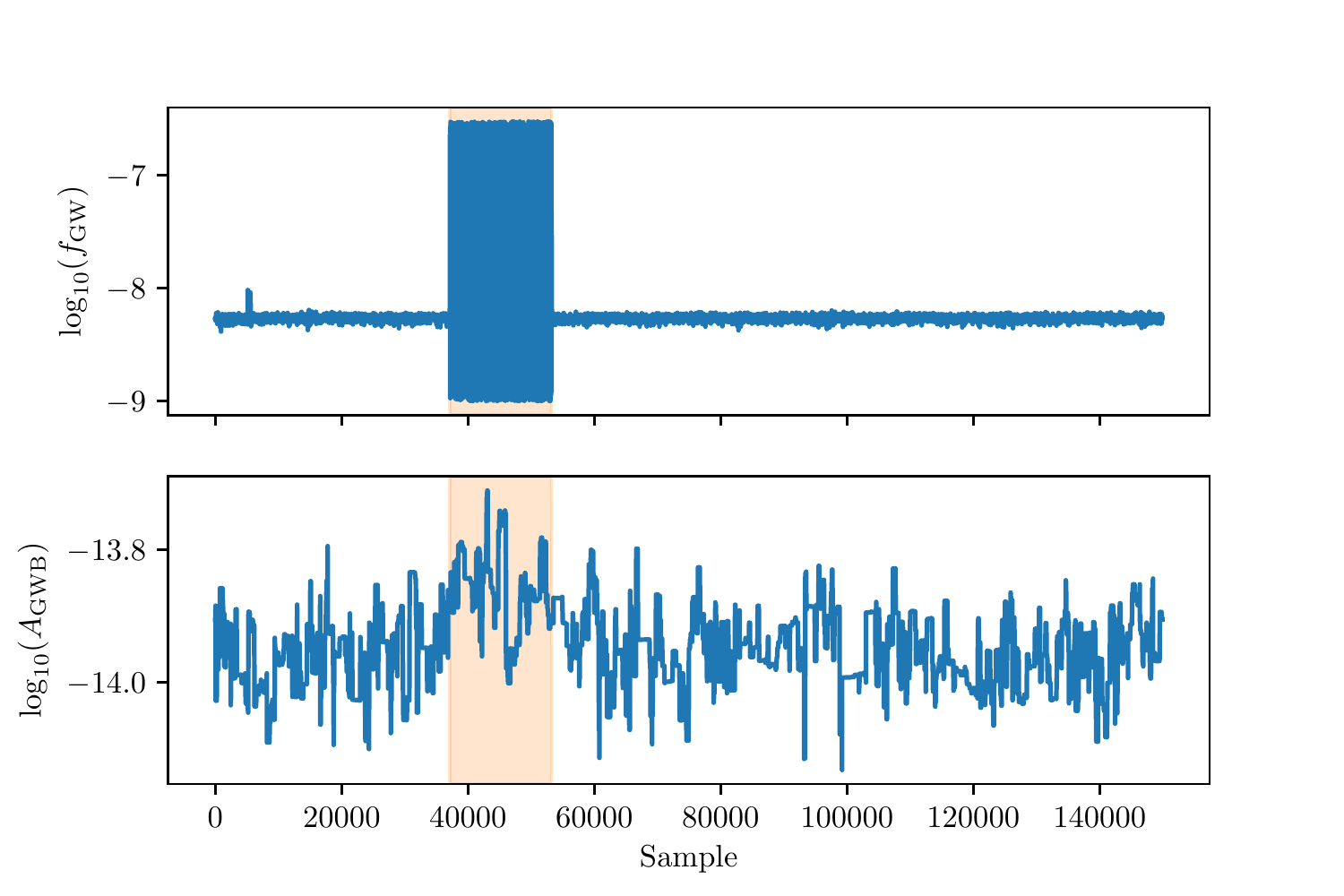}
    \caption{Trace plots for closed dataset 2 showing the frequency of the CW (upper panel) and the GWB amplitude (lower panel) at each sample in the Markov chain.
    It is apparent that the sampler switches modes between finding both the CW and GWB, and finding only a higher amplitude GWB.
    The region where the the CW is not found are highlighted in orange for comparison.}
    \label{fig:closed2_mode_switch}
\end{figure}

In the GWB + CW analysis it appears that the sampler switched between two modes: one with a CW and GWB (presumably the correct solution), and one with a larger amplitude GWB and no CW as shown in Figure \ref{fig:closed2_mode_switch}.
This implies that the chain never truly converged on the correct solution, and more sophisticated sampling methods may be needed. 
This result highlights that in order to make a real combined GWB + CW detection, a vast overhaul of existing methods is necessary, as well as significant testing to prove the reliability of such methods. In the analysis of real PTA data, a GWB will be present when searching for CW signals.
Therefore, it is critical to design methods that are capable of robustly detecting CWs with a GWB going forward.

\input{tabs/closed2_GWB_table.tex}
\input{tabs/closed2_CW_table.tex}

\subsection{Closed Dataset \texttt{g2.d3}}
As with the other closed datasets we analysed \texttt{g2.d3} for a GWB using both the CRN and HD spacial correlation models.
We found no stochastic GWB present in the data, as can be seen in Table \ref{tab:closed3GWB}. 
We therefore deduced that this dataset must contain the two CW signals.

We decided a CW method 1 analysis (see \S\ref{sec:open.g1.d3}) was unfeasible, due to the large number of parameters required.
Instead, we adapted a method 3 type run to simultaneously search for two separate CW sources.
We began by prefitting the WN parameters, including a CW signal in the noise model.
This analysis used only one CW signal, potentially biasing the final WN parameters fixed in the GW analysis.
\par

\begin{figure}
    \centering
    \includegraphics[width = \textwidth]{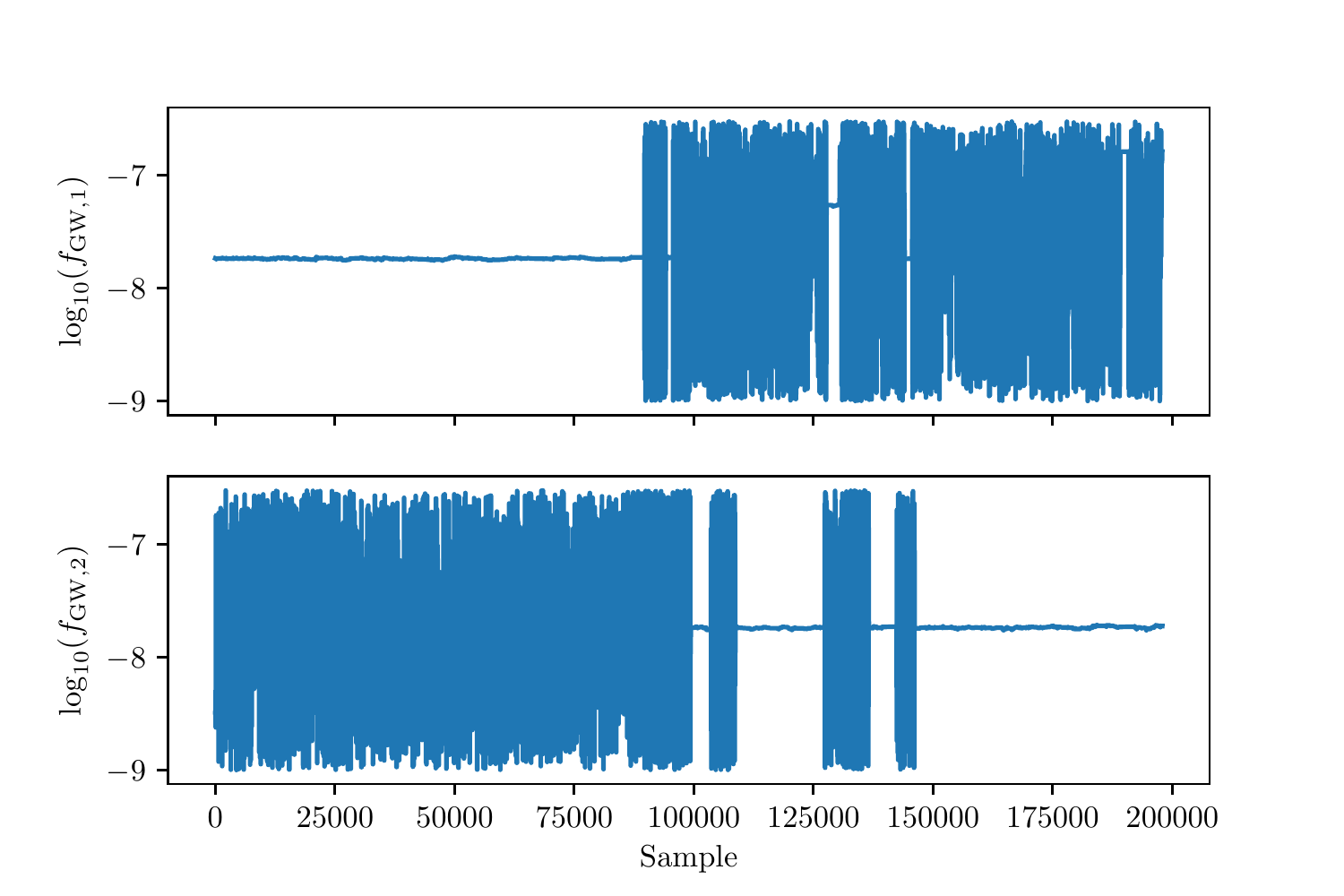}
    \caption{Trace plots for closed dataset 3 showing the frequency of the first CW (upper panel) and the second CW (lower panel) at each sample in the Markov chain.
    It is apparent that the sampler switches modes between finding the CW in the first signal and not the second, and vice versa.
    However, it is clear that the recovered CW is the same in both signals, as the frequency of the GW is the same value.}
    \label{fig:closed3_mode_switch}
\end{figure}

We successfully detected one of the CW signals, and did not detect the second, as summarized in Table~\ref{tab:closed3CW}.
Like the analysis of \texttt{g2.d2}, this analysis also showed signs of mode switching in the posterior distribution.
In this case the one detected signal would alternately appear in one or the other CW model, as is shown in \autoref{fig:closed3_mode_switch}.
The development of more sophisticated sampling methods will be necessary to reliably separate two GW signals.

One interesting complication when searching for two CW signals is that \enterprise sets the pulsar distances as free parameters for each CW signal.
Without significant revisions to the software, the pulsar distance parameters could not be coupled between the two individual CW signals, and were left as independent.
Since the pulsars should have only one distance, this type of analysis clearly needs improvement before it can be carried out on real data.

While not explored in this work, this type of analysis could be implemented by searching for each CW signal one at a time.
This would involve searching for the higher amplitude CW, subtracting it, and then searching for the second, weaker signal.

\input{tabs/closed3_GWB_table.tex}
\input{tabs/closed3_CW_table.tex}

%% file: tabs/closed1_table.tex
\begin{table}[t]
    \scriptsize
    \caption{Detection comparison for \texttt{g2.d1}.
            Bayes factors are for GW and noise vs. noise only,
            and the uncertainties on median are 90\% CIs.
            We report the results of modeling common red noise with no spatial correlations (CRN) and with spatial correlations (HD).
            The last row shows the recovered GWB parameters when a continuous wave is included with the HD model.}
    \begin{center}
    \begin{tabular}{@{} cccc @{}}
        \hline\hline
            run type & $A_{\rm{med}}\; (\times10^{-15})$ & $A_{\rm{mode}}\; (\times10^{-15})$ & $\mathcal{B}$ \\
        \hline
            CRN & $1.9^{+1.7}_{-1.9}$ & $2.0$ & $4.7$ \\
            HD & $2.0^{+1.8}_{-2.0}$ & $2.4$ & $4.2$\\
            \hline
            HD + CW & $11^{+3}_{-2}$ & $12$ & $\infty$ \\
            \hline
    \end{tabular}
    \end{center}
    \label{tab:closed1}
\end{table}

%% file: tabs/closed2_GWB_table.tex
\begin{table}[t]
    \scriptsize
    \caption{Gravitational wave background detection comparison for \texttt{g2.d2}.
            Bayes factors are for GW and noise vs. noise only,
            and the uncertainties on median are 90\% CIs.
            We report the results of modeling common red noise with no spatial correlations (CRN) and with spatial correlations (HD) for detection runs without a continuous gravitational wave (CW).
            The last set of recovered parameters are the GWB values when a CW is included in the model.}
    \begin{center}
    \begin{tabular}{@{} cccc @{}}
        \hline\hline
            run type & $A_{\rm{med}}\; (\times10^{-15})$ & $A_{\rm{mode}}\; (\times10^{-15})$ & $\mathcal{B}$ \\
            \hline
            CRN & $<3.7$ & $2.5$  & $1.3$ \\
            HD & $<3.7$ & $2.4$  & $3.0$ \\
            \hline
            HD + CW (CW + GWB Mode) & $11^{+3}_{-2}$ & $11$ & $\infty$ \\
            HD + CW (GWB Only Mode) & $14^{+4}_{-3}$ & $13$ & $\infty$ \\
            \hline
    \end{tabular}
    \end{center}
    \label{tab:closed2GWB}
\end{table}

%% file: tabs/closed2_CW_table.tex
\begin{table*}[t]
    \scriptsize
    \caption{Continuous gravitational wave detection parameters for the continuous wave present in \texttt{g2.d2}.
            }
    \begin{center}
    \begin{tabular}{@{} ccccccccc @{}}
        \hline\hline
            Mode &
            $f_{\rm{GW}}$ & $\mathcal{M}_{c}$ & $h$ & $\phi_{\rm{GW}}$ & $\theta_{\rm{GW}}$ & $\iota$ & $D$ & $\mathcal{B}$\\
            &
            ($\times 10^{-9}$ Hz) & ($\times 10^{9}$ M$_{\odot}$) & ($\times 10^{-14}$) & (rad) & (rad) & (rad) & (Mpc) &  \\
            \hline
            CW + GWB Mode & $5.4^{+0.3}_{-0.5}$ & $0.2^{+4.9}_{-0.2}$ & $8^{+4}_{-5}$ & $4.0^{+0.2}_{-0.5}$ & $1^{+2}_{-1}$ & $1.6^{+0.8}_{-0.4}$ & $0.3^{+49.1}_{-0.3}$ & $\infty$ \\
            GWB Only Mode & -- & -- & $ <$ 5.98 & -- & -- & -- & -- & $ 1.2$ \\
            \hline
    \end{tabular}
    \end{center}
    \label{tab:closed2CW}
\end{table*}

%% file: tabs/closed3_GWB_table.tex
\begin{table}[t]
    \scriptsize
    \caption{Gravitational wave background detection comparison for \texttt{g2.d3}.
            Bayes factors are vs. noise only.
            Uncertainty on median gives 90\% CI. 
            We report the results of modeling common red noise with no spatial correlations (CRN) and with spatial  correlations (HD) for detection runs without a continuous gravitational wave (CW).
            }
    \begin{center}
    \begin{tabular}{@{} cccc @{}}
        \hline\hline
            run type & $A_{\rm{med}}\; (\times10^{-15})$ & $A_{\rm{mode}}\; (\times10^{-15})$ & $\mathcal{B}$ \\
            \hline
            CRN & $<0.3$ & $0.1$  & $0.5$ \\
            HD & $<0.3$ & $0.1$  & $0.5$ \\
            \hline
    \end{tabular}
    \end{center}
    \label{tab:closed3GWB}
\end{table}

%% file: tabs/closed3_CW_table.tex
\begin{table*}[t]
    \scriptsize
    \caption{Continuous gravitational wave detection parameters for both continuous waves present in \texttt{g2.d3}.
            }
    \begin{center}
    \begin{tabular}{@{} ccccccccc @{}}
        \hline\hline
           Signal & $f_{\rm{GW}}$ & $\mathcal{M}_{c}$ & $h$ & $\phi_{\rm{GW}}$ & $\theta_{\rm{GW}}$ & $\iota$ & $D$ & $\mathcal{B}$\\
           &
            ($\times 10^{-9}$ Hz) & ($\times 10^{9}$ M$_{\odot}$) & ($\times 10^{-14}$) & (rad) & (rad) & (rad) & (Mpc) &  \\
            \hline
           CW 1&  $18.3^{+0.4}_{-0.5}$ & $0.9^{+2.1}_{-0.9}$ & $1.4^{+2.1}_{-0.7}$ & $1.0^{+1.1}_{-0.8}$ & $2^{+3}_{-1}$ & $1.0^{+0.5}_{-0.7}$ & $30^{+340}_{-30}$ & $\infty$ \\
           CW 2 & 
            -- & -- & $ < 1.068 $ & -- & -- & -- & -- & $1.0$ \\
            \hline
    \end{tabular}
    \end{center}
    \label{tab:closed3CW}
\end{table*}

%% file: sec_discuss.tex
For the most part, our work on MDC2 followed the standard GW data analysis procedures of NANOGrav.
Participating in MDC2 allowed us to rigorously test that methodology, which is an important endeavour to undertake.
During the MDC we verified the overall procedure, illuminated some known issues (using incorrect models, sampling high dimensional spaces), and uncovered one potential shortcoming of the standard CW analysis (prefitting WN).
\par

Using the two open datasets which contained only a GWB (\texttt{g1.d1} and \texttt{g1.d2}), we were unable to claim a confident detection in several cases.
For \texttt{g1.d1} we found that using a fixed SSE led to a confident detection, whereas using \bayesephem did not.
We were unable to detect the GWB in \texttt{g1.d2}, although our upper limits that are consistent with the injected background.
\par

Our confident detection of the weaker signal in \texttt{g1.d1} and non-detection of the louder signal in \texttt{g1.d2} is certainly due to the differences in injected noise.
Dataset \texttt{g1.d1} contained WN only, while some pulsars in \texttt{g1.d2} had intrinsic RN as well.
The addition of RN greatly affects detection prospects.
This analysis underscores the fact that detection strategies based on WN-only models will be wildly optimistic.
For a more detailed look on how WN and RN affect PTA sensitivity see \cite{hasasia}.
\par

For \texttt{g1.d3}, which contained a single CW, we confidently detected the CW signal in all cases.
A spurious GWB was detected in the case where WN parameters were prefit using single pulsar analyses containing WN, RN, and no CW, despite the lack of a GWB injection in the dataset.
We suspect that mismodeling during the noise analyses led to this erroneous signal.
The true source of this signal remains unclear and warrants further investigation beyond the scope of this work.
This result should be seen as a warning that great care must be taken when selecting data analysis models.
\par

We concluded that the closed dataset \texttt{g2.d1} contained no CW source, thus only a GWB.
We did not confidently detect the GWB signal.
We found when simultaneously searching for both a GWB and a CW, we again found a spurious signal.
We confidently detected a GWB with amplitude much larger than injected.
The reason for this warrants further investigation beyond the scope of this work, but we suspect that the unsupported CW model introduced residual power, which was compensted by the GWB model.
Again mismodeling is the problem.
\par

We confidently detected a single CW source in closed dataset \texttt{g2.d2} and found weak evidence for a GWB.
Finally, We concluded that dataset \texttt{g2.d3} contained two CW sources, even though we were only able to detect one of them.

In addition to the standard analysis techniques used by NANOGrav, we have expanded the PTA data analysis methods, particularly when it comes to CW searches.
Previously published studies have been limited to searches for GWB or searches for single CWs \citep[e.g.,][]{ng11.gwb, ng11.cw}.
This MDC allowed us to perform a search for both a GWB and a CW simultaneously.
The implementation of this search in \enterprise is straightforward and has been done by others exploratorily.
To our knowledge, this is the first time that results from such a search have been presented, even on simulated data.
Searching for multiple CW sources simultaneously is a second straightforward implementation in \enterprise, but admits practical sampling challenges.
Finally, in our work we searched directly over the GW frequency as a parameter in our CW analyses. 
This differs from \citet{ng11.cw} who performed multiple independent searches, each over a fixed frequency bin.
The direct frequency sampling methods we employ are again straightforward to implement in \enterprise, but present practical sampling challenges.
The methodology we use was developed for \cite{wittinprep}.
\par

We were able to achieve success with these advancements in CW searches using \ptmcmc.
However, the MCMC chains in our CW analyses required long burn in times and have long autocorrelation lengths.
In order to achieve our modest level of sampling success, we had to manually tune \ptmcmc.
In particular we supplied sampling ``groups'', subsets of parameters we believed to be correlated a priori.
\ptmcmc uses these groups to propose jumps in multiple dimensions simultaneously.
We also tuned the proposal distributions so that some parameters were jumped more often than others.
\par

Through the course of participating in MDC2 we had to change our method for prefitting WN parameters.
In the past when prefitting WN, single pulsars were analyzed modeling only WN and RN.
The single RN power-law is assumed to be able to handle all excess low frequency features including intrinsic pulsar RN, GWB, and even any strong CW signals.
We found that, despite standard practices presented in \citet{ng11.gwb, ng11.cw}, in order to accurately and efficiently estimate the WN parameters in the presence of detectable CWs, it is critical to include the CW waveform in the single pulsar noise analyses.
This fact points to a possible further complication: if both intrinsic pulsar noise and GWB are present with significantly different spectral slopes, a simple power law RN model may not be sufficient for any noise analyses.
This fact may have contributed to our poor recovery of WN parameters for \texttt{g1.d2}.
After a GWB is detected and PTAs focus on characterizing the background, we will need to properly handle the GWB in the noise prefitting procedure so as not to bias WN parameters.
\par

For each analysis, the MCMC jobs took of order a few days to a week, depending on the type of modeling employed, with a maximum of 12 CPU cores used.
This MDC had a much smaller data volume than a typical real PTA dataset.
Each MDC2 dataset contained 33 pulsars each with 183 TOAs, for a total of 6,039 TOAs.
The simulated datasets contain one TOA per observation, whereas real PTA data can contain multiple TOAs for each frequency band of a single observation.
Current PTA datasets also contain many more pulsars, for example the second data release of the IPTA contains data for 65 pulsars.
As a comparison the NANOGrav 11-year data release contains $\mathcal{O}(10^5)$~TOAs total \cite{ng11.data}.
The next generation of PTA datasets could approach $\mathcal{O}(10^6)$ TOAs.
Given the increasing size of datasets, the GW analysis computation time will be significantly lengthened without algorithmic advances.
\par

The use of new TOA generation methods could mitigate this.  For example wideband pulsar timing combines data from multiple radio frequency bands to produce one single TOA \citep{pdr2014, Pennucci2019}.
Methods like this can dramatically reduce the number of TOAs in existing datasets and slow their growth into the future.
\par

Given the difficulties in executing our sampling methods (e.g., \S\ref{sec:closed.g2.d2}), along with the time taken per analysis, we find it imperative that new approaches be developed.
As discussed in \S\ref{sec:analysis}, \enterprise currently uses an MCMC sampler that is capable of parallel tempering, but that component was not used for these analyses.
Parallel tempering uses a set of $N$ different chains that explore the parameter space at different ``temperatures'', effectively rescaling the likelihood surface.
The ``hotter'' chains analyze on a flattened version of the surface and can more freely move between local maxima.
The chains communicate with each other, ``hotter'' passing information to the ``colder'' ones \citep{Geyer1991}.
Simply using \ptmcmc's full capabilities could improve our analyses.
\par

Using different MCMC algorithms could also improve sampling efficiency.
In particular Hamiltonian Monte Carlo sampling (HMC) is well suited to high dimensional parameter spaces \citep{hmc}.
HMC has been implemented in widely used software like \texttt{PyMC3} \citep{pymc3}.
One of the strengths of the \enterprise framework is that it generates the likelihood and prior functions, but allows the user to do with these whatever they choose.
There is a great opportunity to leverage well maintained computational statistics software libraries to improved PTA data analysis with \enterprise.
\par

In addition to improving sampling efficiency, speeding up the likelihood computation could improve the analysis by allowing more posterior samples to be generated using less clock time.
The PTA likelihood is limited by slow linear algebra calculations, specifically inverting large matrices.
These linear algebra routines are suitable for GPU computation, as each element of the output matrix can be calculated independently of of the others.
The boom of machine learning research has lead to several software libraries offload linear algebra to GPUs, e.g. \texttt{TensorFlow} \citep{tensorflow}.
There are likely large gains to be made in computation by using GPUs to perform the linear algebra calculations in \enterprise.

%% file: sec_appendix.tex
\section{Ephemeris Oddities}
\label{sec:appendix.eph}

\begin{figure}
    \centering
    \subfloat[]{
        \includegraphics[width = 0.48\linewidth]{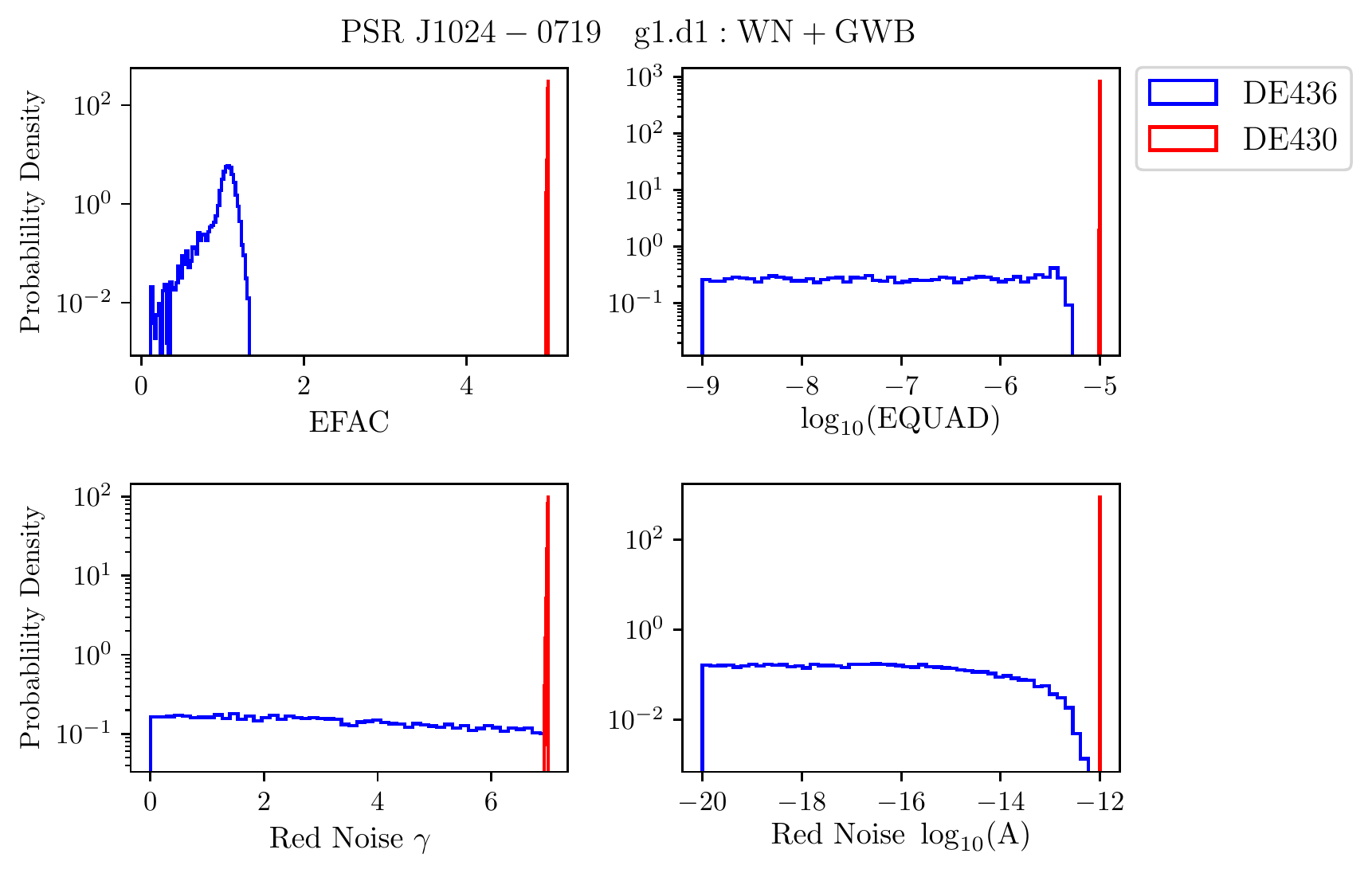}
        \label{fig:g1.d1_J1024}
    }
    \subfloat[]{
        \includegraphics[width = 0.48\linewidth]{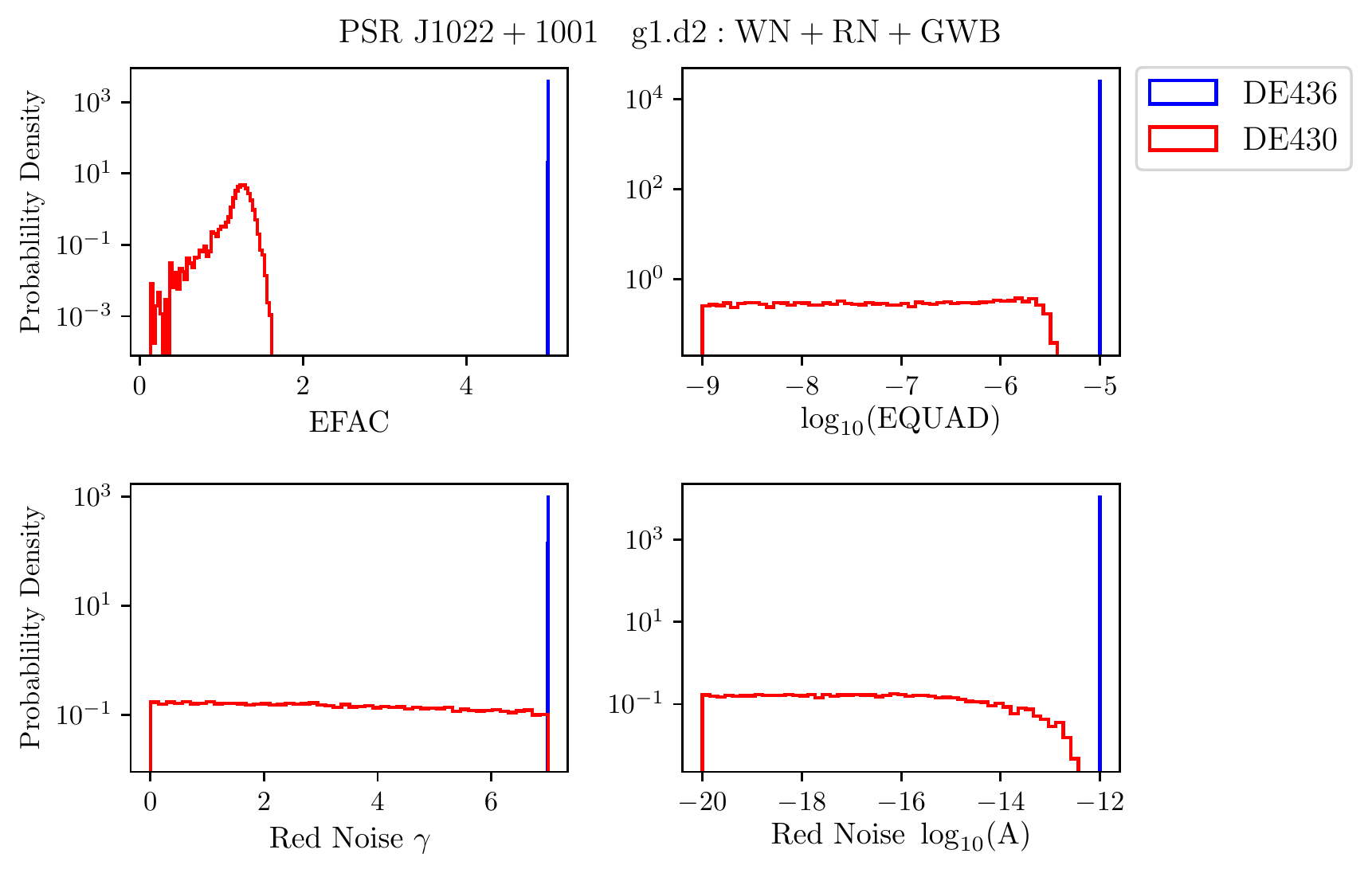}
        \label{fig:g1.d2_J1022}
    }
    \newline
    \subfloat[]{
        \includegraphics[width = 0.48\linewidth]{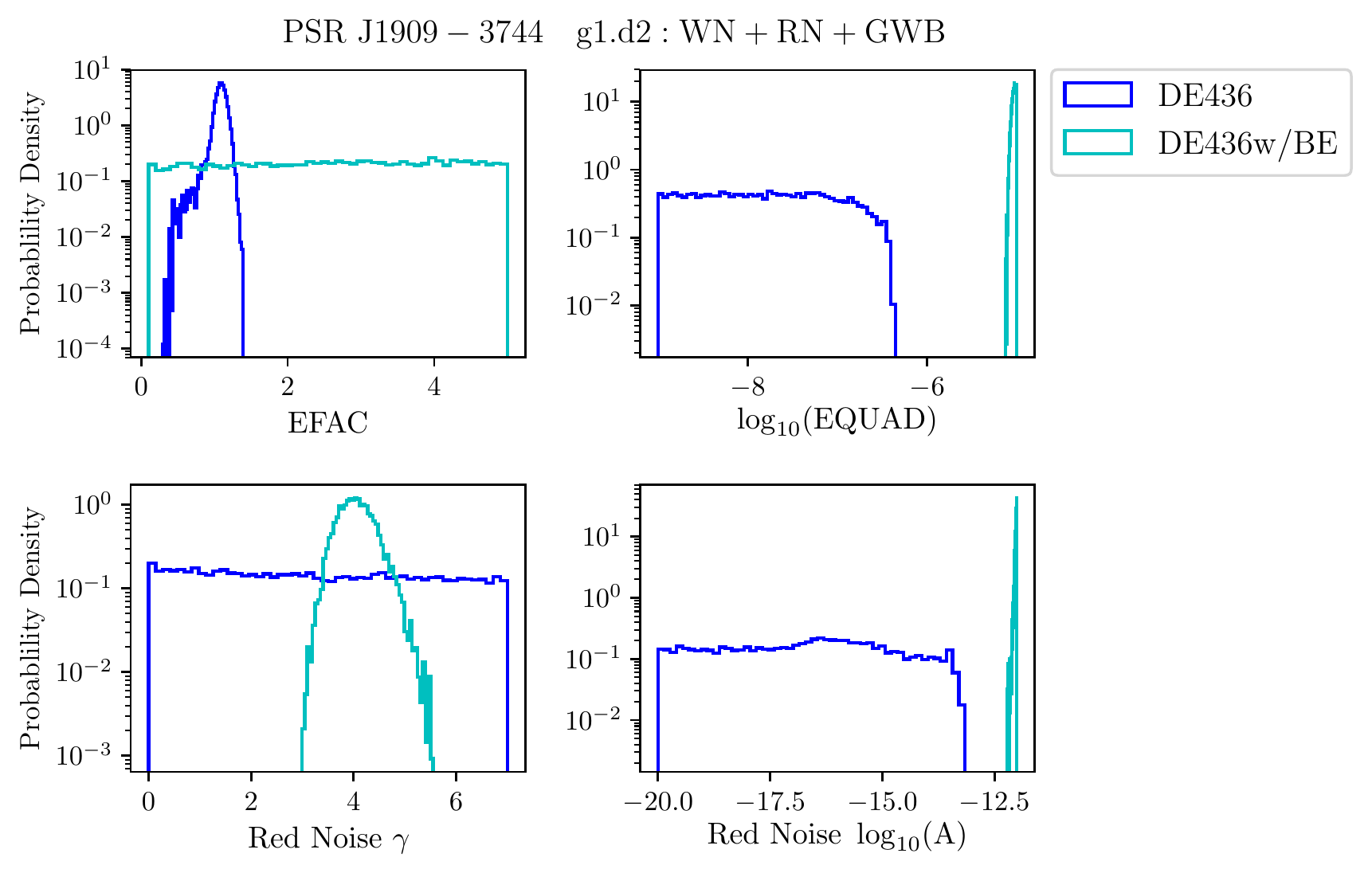}
        \label{fig:g1.d2_J1909}
    }
    \caption{Posteriors of both white and red noise for: (a) PSR~J1024$-$0719 using two different ephemerides, DE430 and DE436, for dataset \texttt{g1.d1}, (b) PSR~J1022$+$1001 using two different ephemerides, DE430 and DE436, for dataset \texttt{g1.d2}, and (c) PSR~J1909$-$3744 for \texttt{g1.d2} using DE436 with and without \bayesephem. (a) and (b) show unexpected differences for the parameters obtained using DE430 when compared to DE436, and (c) likewise obtain significantly different parameters when using DE436 with and without \bayesephem.
    }
    \label{fig:weird_pulsar_posteriors}
\end{figure}

There are a number of differences between the different ephemeris runs for \texttt{g1.d1} and \texttt{g1.d2}.
For some pulsars there were large differences between both the white and red noise parameters when switching between DE436 and DE430 for \texttt{g1.d1} and additionally between DE436 with and without \bayesephem for \texttt{g1.d2}.
For \texttt{g1.d1} we note that using either ephemeris results in a detection of the GWB. 
This could be because the error induced from using the wrong ephemeris is likely at the same level as the GWB.
In particular for \texttt{g1.d1}, there was a significant difference in both the white and red noise parameters for PSR~J1024$-$0719.
These differences are shown in Figure \ref{fig:g1.d1_J1024}.
All runs that included \bayesephem showed posteriors that agreed well with what was sampled using DE436 without \bayesephem.

Additionally, we found that for \texttt{g1.d2} there were a number of pulsars that showed significant differences in both the white and red noise parameter posteriors when switching between DE436 with and without \bayesephem.
One pulsar, PSR~J1022$+$1001, showed differences between DE430 and DE436 as in \texttt{g1.d1}.
These differences are shown in Figure \ref{fig:weird_pulsar_posteriors}.

A number of other pulsars showed different white and red noise parameters for different ephemerides.
PSR~J1022$+$1001 is the only pulsar that shows differences between DE436 and DE430 in\texttt{g1.d2}.
For this pulsar the posteriors are similar to what is shown for PSR~J1024$-$0719 in \texttt{g1.d1}.
PSR~J1909$-$3744 is one of four pulsars that shows different red and white noise parameters when using DE436 with and without \bayesephem.
The red and white noise posteriors for PSR~J1909$-$3744 are shown in Figure \ref{fig:g1.d2_J1909}.

In addition, PSRs~J1024$-$0719, J1918$-$0642, and J1939$+$2134 showed similar white and red noise deviations when using DE436 with \bayesephem to those shown in Figure \ref{fig:weird_pulsar_posteriors}.

\section{Continuous Wave Parameter Estimation}
\label{sec:appendix.cw}

\begin{figure}
    \centering
    \includegraphics[width=\textwidth]{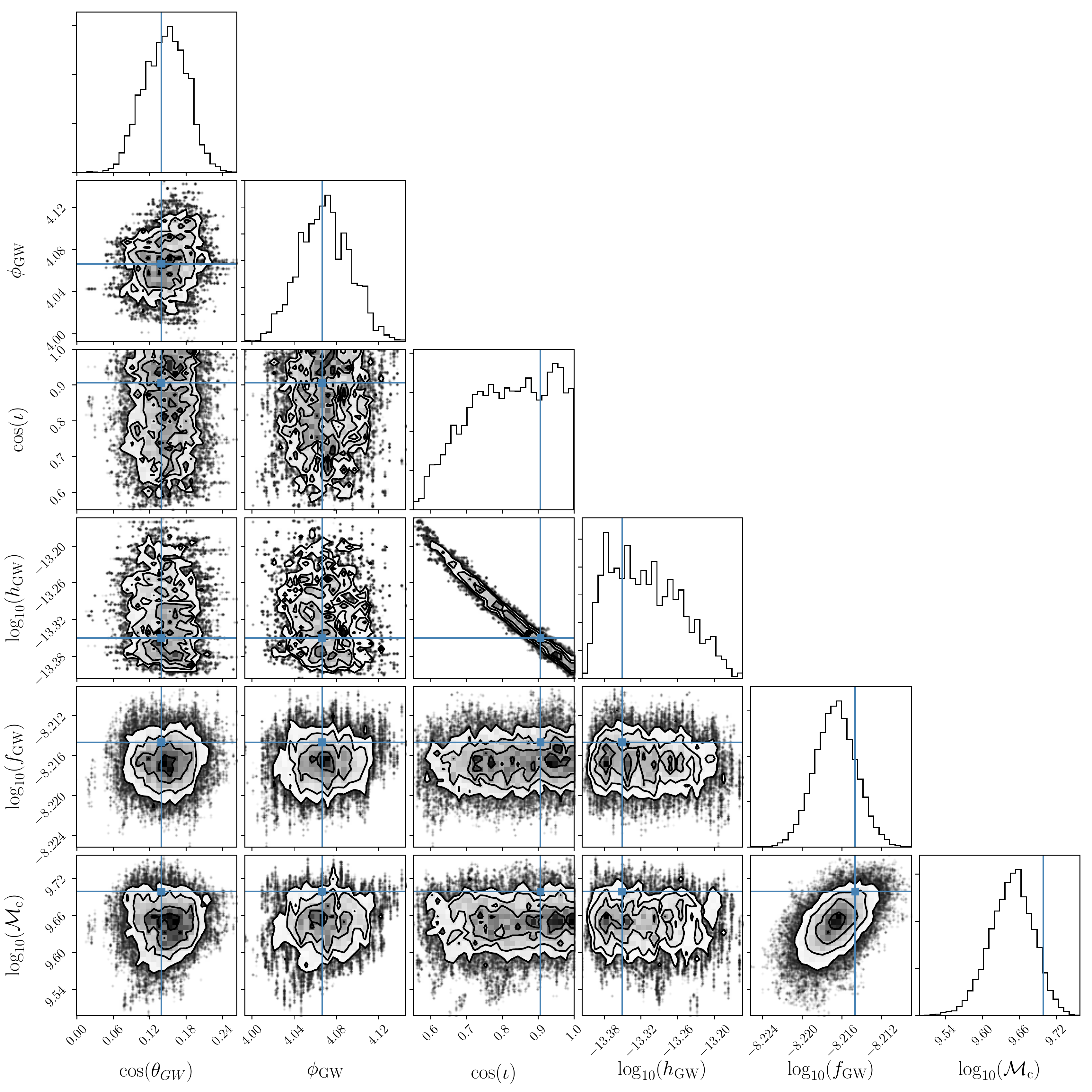}
    \caption{CW parameter posteriors for \texttt{g1.d3} using Method 1. Here $f_{\rm{GW}}$ is the GW frequency, $\mathcal{M}_c$ is the chirp mass of the binary, $(h_{\rm{GW}})$ is the characteristic strain, $\theta_{\rm{GW}}$ and $\phi_{\rm{GW}}$ are the source position on the sky, and $\iota$ is the orbital inclination of the binary.}
    \label{fig:g1.d3_corner_m1}
\end{figure}

\begin{figure}
    \centering
    \includegraphics[width=\textwidth]{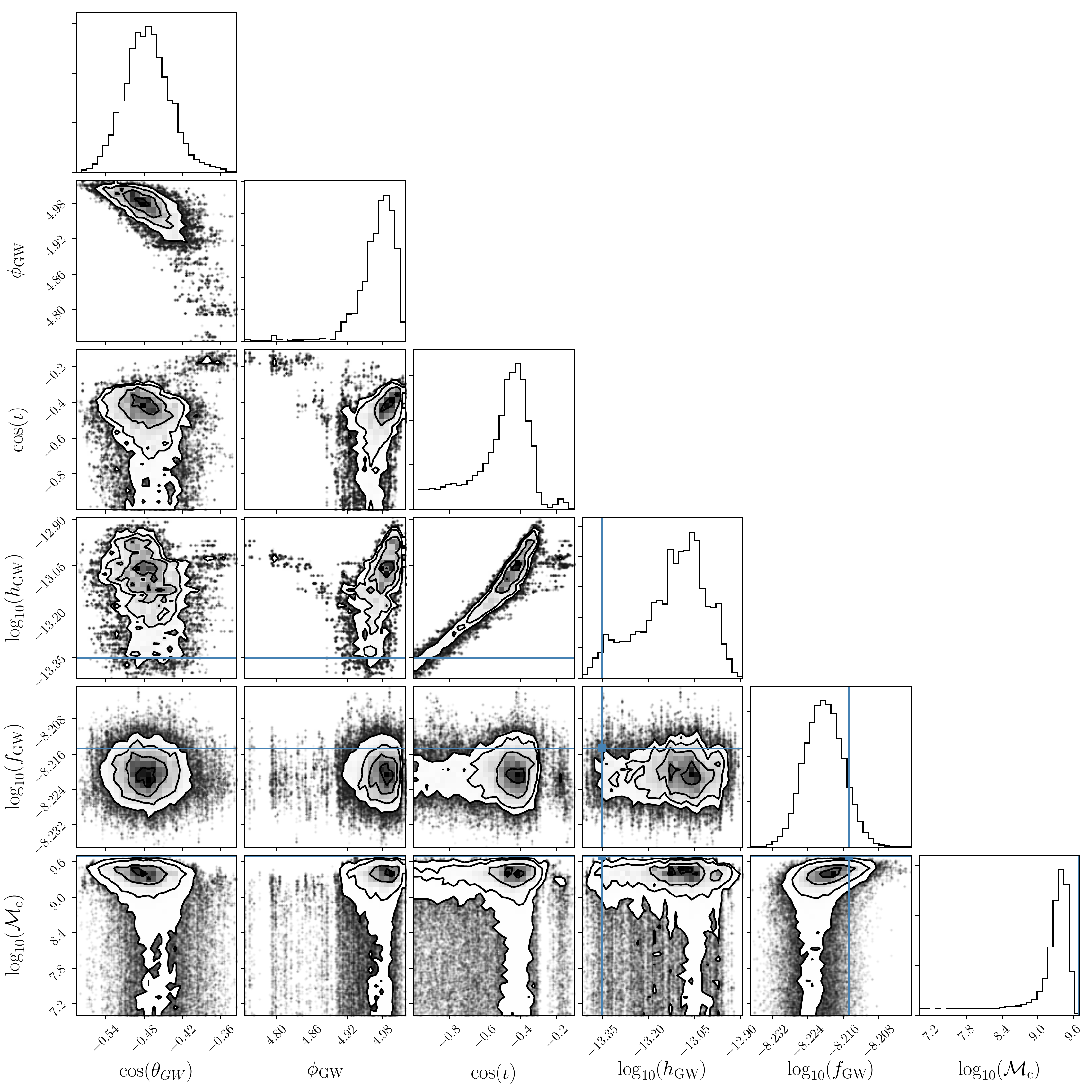}
    \caption{CW parameter posteriors for \texttt{g1.d3} using Method 2. Here $f_{\rm{GW}}$ is the GW frequency, $\mathcal{M}_c$ is the chirp mass of the binary, $(h_{\rm{GW}})$ is the characteristic strain, $\theta_{\rm{GW}}$ and $\phi_{\rm{GW}}$ are the source position on the sky, and $\iota$ is the orbital inclination of the binary.
    }
    \label{fig:g1.d3_corner_m2}
\end{figure}

\begin{figure}
    \centering
    \includegraphics[width=\textwidth]{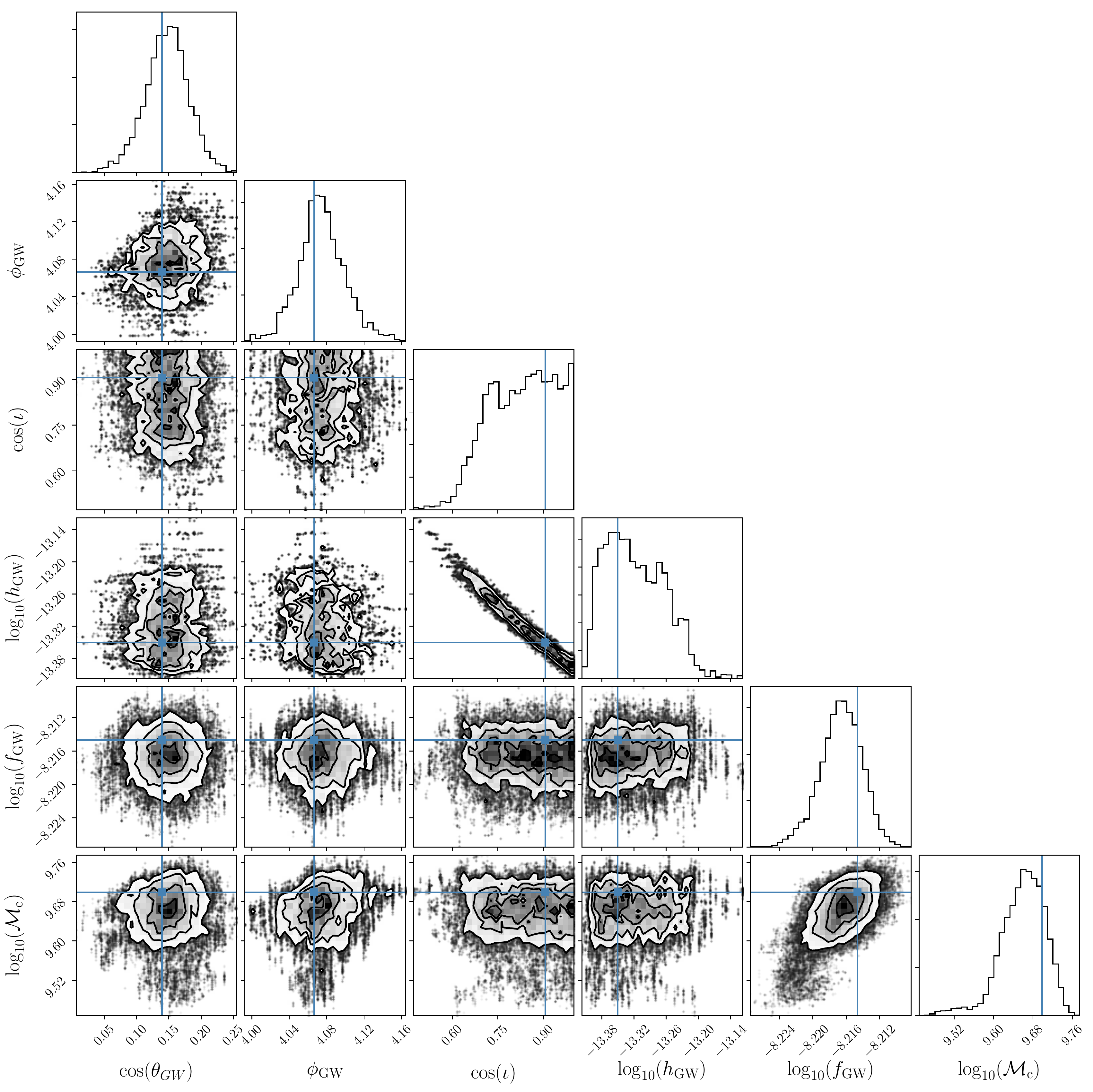}
    \caption{CW parameter posteriors for \texttt{g1.d3} using Method 3. Here $f_{\rm{GW}}$ is the GW frequency, $\mathcal{M}_c$ is the chirp mass of the binary, $(h_{\rm{GW}})$ is the characteristic strain, $\theta_{\rm{GW}}$ and $\phi_{\rm{GW}}$ are the source position on the sky, and $\iota$ is the orbital inclination of the binary.}
    \label{fig:g1.d3_corner_m3}
\end{figure}

In \S\ref{sec:open.g1.d3} we analyzed dataset \texttt{g1.d3} with three different methods.
Here we present parameter posteriors for each of these methods in Figures \ref{fig:g1.d3_corner_m1}, \ref{fig:g1.d3_corner_m2}, and \ref{fig:g1.d3_corner_m3}.

%% file: biblio.tex
\newcommand{\apj}{\textit{Astrophys.~J.}}
\newcommand{\apjl}{\textit{Astrophys.~J.~Let.}}
\newcommand{\apjs}{\textit{Astrophys.~J.~Sup.}}
\newcommand{\mnras}{\textit{MNRAS}}
\newcommand{\prd}{\textit{Phys.~Rev.~D}}
\newcommand{\cqg}{\textit{Class.~Quant.~Grav.}}

\newcommand{\newblock}{} 